\documentclass[fleqn,usenatbib,usegraphicx]{mn2e}

\usepackage{aas_macros}
\usepackage{amsmath,amssymb}
\usepackage[subrefformat=parens]{subfig}
\usepackage{url}

\title{Sinking of a magnetically confined mountain on an accreting neutron star}

\author[K. Wette, M. Vigelius and A. Melatos]{
K. Wette$^{1}$\thanks{E-mail: karl.wette@anu.edu.au}, M. Vigelius$^{2}$ and A. Melatos$^{2}$ \\
$^{1}$ Centre for Gravitational Physics, Australian National University, Canberra, ACT 0200, Australia \\
$^{2}$ School of Physics, University of Melbourne, Parkville, VIC 3010, Australia}

\newcommand{\Msun}{M_{\sun}}
\newcommand{\cs}{c_\mathrm{s}}
\newcommand{\Mstar}{M_{\star}}
\newcommand{\Rstar}{R_{\star}}
\newcommand{\Bstar}{B_{\star}}
\newcommand{\psistar}{\psi_{\star}}
\newcommand{\psia}{\psi_\mathrm{a}}
\newcommand{\Mc}{M_\mathrm{c}}
\newcommand{\Ma}{M_\mathrm{a}}
\newcommand{\Mbase}{M_\mathrm{base}}
\newcommand{\Matmos}{M_\mathrm{atm}}
\newcommand{\rmin}{r_\mathrm{min}}
\newcommand{\rmax}{r_\mathrm{max}}
\newcommand{\rhosurf}{\rho_{\Rstar}}
\newcommand{\rinj}{r_\mathrm{inj}}
\newcommand{\drinj}{\delta r_\mathrm{inj}}
\newcommand{\Madot}{\dot{\Ma}}
\newcommand{\Tinj}{T_\mathrm{a}}
\newcommand{\tmax}{t_\mathrm{max}}
\newcommand{\vinj}{v_\mathrm{inj}}
\newcommand{\Xmnt}{\mathrm{X}_\mathrm{a}}
\newcommand{\talfv}{t_\text{Alfv\'en}}
\newcommand{\valfv}{v_\text{Alfv\'en}}
\newcommand{\vb}{\mathbf{v}}
\newcommand{\Bb}{\mathbf{B}}
\newcommand{\Ek}{E_\mathrm{k}}
\newcommand{\Em}{E_\mathrm{m}}
\newcommand{\Izz}{I_\mathrm{zz}}

\newcommand{\Mi}{1}
\newcommand{\Mii}{10}
\newcommand{\Miii}{10^2}
\newcommand{\Miiii}{10^3}
\newcommand{\probh}{\mathcal{H}}
\newcommand{\probhi}{\mathcal{H}(\Mi)}
\newcommand{\probhii}{\mathcal{H}(\Mii)}
\newcommand{\probhiii}{\mathcal{H}(\Miii)}
\newcommand{\probhiiii}{\mathcal{H}(\Miiii)}
\newcommand{\probs}{\mathcal{S}(\rmin)}
\newcommand{\probsi}{\mathcal{S}(\rmin,\Mi)}
\newcommand{\probsii}{\mathcal{S}(\rmin,\Mii)}
\newcommand{\probsiii}{\mathcal{S}(\rmin,\Miii)}
\newcommand{\probsiiii}{\mathcal{S}(\rmin,\Miiii)}
\newcommand{\probss}{\mathcal{S}(\Rstar)}
\newcommand{\probssi}{\mathcal{S}(\Rstar,\Mi)}
\newcommand{\probssii}{\mathcal{S}(\Rstar,\Mii)}
\newcommand{\probssiii}{\mathcal{S}(\Rstar,\Miii)}
\newcommand{\probssiiii}{\mathcal{S}(\Rstar,\Miiii)}

\begin{document}

\maketitle

\begin{abstract}
We perform ideal-magnetohydrodynamic axisymmetric simulations of magnetically confined mountains on an accreting neutron star, with masses $\lesssim 0.12 \Msun$.
We consider two scenarios, in which the mountain sits atop a hard surface or sinks into a soft, fluid base.
We find that the ellipticity of the star, due to a mountain grown on a hard surface, approaches $\sim 2 \times 10^{-4}$ for accreted masses $\gtrsim 1.2 \times 10^{-3} \Msun$, and that sinking reduces the ellipticity by between 25\% and 60\%.
The consequences for gravitational radiation from low-mass x-ray binaries are discussed.
\end{abstract}


\section{Introduction}

The magnetic dipole moment $\mu$ of a neutron star is observed to diminish in the long term as the star accretes \citep{Taam-vdHeuvel-1986,vdHeuvel-Bitzaraki-1995}, although \citet{Wijers-1997} argued that $\mu$ may also be a function of parameters other than the accreted mass $\Ma$.
The $\mu$--$\Ma$ correlation has been ascribed to a number of physical mechanisms \citep{Melatos-Phinney-2001,Cumming-2005}.
First, the magnetic field may be dissipated in the stellar crust by Ohmic decay, accelerated by heating as the accreted plasma impacts upon the star \citep{Konar-Bhattacharya-1997,Urpin-etal-1998,Brown-Bildsten-1998,Cumming-etal-2004}.
Second, magnetic flux tubes may be dragged from the superconducting core by the outward motion of superfluid vortices, as the star spins down \citep{Srinivasan-etal-1990,Ruderman-etal-1998,Konar-Bhattacharya-1999,Konen-Geppert-2001}.
Third, the magnetic field may be screened by accretion-induced currents within the crust \citep{BisnovatyiKogan-Komberg-1974,Blondin-Freese-1986,Lovelace-etal-2005}.
In particular, the field may be \emph{buried} under a mountain of accreted plasma channelled onto the magnetic poles.
When $\Ma$ is large enough, the mountain spreads laterally, transporting the polar magnetic flux towards the equator \citep{Hameury-etal-1983,Romani-1990,Brown-Bildsten-1998,Cumming-etal-01,Melatos-Phinney-2001,
Choudhuri-Konar-2002,PM04,Zhang-Kojima-2006,PM07,VM08,VM09R}.

\citet{PM04} computed the unique sequence of self-consistent, ideal-magneto\-hydro\-dynamic (ideal-MHD) equilibria that describes the formation of a polar mountain by magnetic burial as a function of $\Ma$.
They found that the accreted mountain is confined by the equatorially compressed magnetic field, which was unaccounted for in previous calculations, and that $10^{-5}\Msun$ must be accreted to lower $\mu$ by 10\%.
Surprisingly, mountains are stable with respect to axisymmetric ideal-MHD perturbations; they oscillate globally in a superposition of acoustic and Alfv\'en modes but remain intact due to magnetic line-tying at the stellar surface \citep{PM07}.
The same equilibria are susceptible to nonaxisymmetric, Parker-like instabilities (specifically the gravitationally driven, undular sub-mode), but the instability preserves a polar mountain when it saturates, despite reducing the mass ellipticity by $\sim 30\%$ \citep{VM08}.
Recently, \citet{VM09R} considered resistive effects.
They found that the mountain does not relax appreciably for realistic resistivities over the lifetime of a low- or high-mass X-ray binary, either by global diffusion, resistive g-mode instabilities, or reconnection in the equatorial magnetic belt.
The Hall drift, which exerts a destabilising influence in isolated neutron stars \citep[see, e.g.,][]{Rheinhardt-Geppert-2002}, is unlikely to be important in accreting neutron stars due to crustal impurities \citep{Cumming-etal-2004,Cumming-2005}.

The investigations outlined in the previous paragraph suffer from two limitations.
First, the mountain is assumed to rest upon a rigid surface.
Under this assumption, the accreting plasma cannot sink into the stellar crust.
This is unrealistic.
During magnetic burial, frozen-in magnetic flux is redistributed slowly within the neutron star by the accreted plasma, as it sinks beneath the surface and spreads laterally.
\citet{Choudhuri-Konar-2002} showed that the time-scale and end state of burial are tied to these slow interior motions.
Second, the accreted plasma is assumed to satisfy an isothermal equation of state.
This is an accurate model only for neutron stars with low accretion rates $\Madot \lesssim 10^{-10} \Msun \mathrm{yr}^{-1}$; the thermodynamics of neutron stars accreting near the Eddington limit ($\sim 10^{-8} \Msun \mathrm{yr}^{-1}$) is more complicated, with a depth-dependent adiabatic index \citep{Brown-Bildsten-1998,Brown-2000}.
The equation of state affects the growth rate of Parker-like instabilities \citep{Kosinski-Hanasz-2006}.

In this paper, we seek to overcome the first limitation.
In section~\ref{sec:method}, we present a new method of computationally simulating the growth of a magnetic mountain with $\Ma \lesssim 0.1 \Msun$.
In section~\ref{sec:compare}, we compare the structure of mountains grown on hard and soft surfaces to evaluate the role of sinking.
In section~\ref{sec:ellip}, the resulting mass quadrupole moment is evaluated as a function of $\Ma$ for hard and soft surfaces.
A comparison with the results of \citet{Choudhuri-Konar-2002}, and the implications for gravitational wave emission from rapidly rotating accretors (e.g. low-mass X-ray binaries), are discussed in section~\ref{sec:discuss}.

\section{Growing a realistically sized mountain by injection}\label{sec:method}

In order to investigate how a magnetically confined mountain sinks into the stellar crust, we need a numerical method capable of building a stable mountain, with a realistic $\Ma$, on top of a fluid base.
The approach we take builds upon previous work by \citet{PM04,PM07} and \citet{VM08,VM09GW}.
Here, as a service to the reader, we briefly recapitulate the physical arguments and key results from these previous papers, with references to the relevant sections and equations.

In \citet{PM04}, axisymmetric magnetic mountain equilibria are computed by solving an elliptic partial differential equation: the Grad-Shadranov equation describing hydromagnetic force balance in axisymmetric geometry [\citet{PM04}, section~2.1 and equation~(12)].
The calculation ensures that the mass-magnetic flux distribution $\partial M / \partial \psi$ is treated self-consistently: the final $\partial M / \partial \psi$ is equal to the initial $\partial M / \partial \psi$ together with the mass-flux distribution of the accreted matter, $\partial \Ma / \partial \psi$, which is characterised by the parameter $\psia$ [\citet{PM04}, section~2.2 and equation~(13)].
In the limit of small $\Ma$, the final equilibrium flux solution is characterised by the ratio $\Ma / \Mc$, where the characteristic mass $\Mc \propto \Mstar \Rstar^2 \Bstar^2$ is the accreted mass required to halve $\mu$ [\citet{PM04}, section~3.2 and equation~(30); \citet{PM07}, section~2.2 and equation~(3)].
The characteristic mass contains the dependence of the equilibrium solution on the parameters of the neutron star, in particular the magnetic field strength $\Bstar$.
The Grad-Shafranov equilibria are computed using an iterative numerical solver [\citet{PM04}, section~3.3]; this approach only converges numerically for low accreted masses $\Ma \le \Mc \approx 10^{-4} \Msun$ [\citet{PM04}, section~3.4], and it fails to accommodate a fluid interior within its fixed-boundary framework.\footnote{
In \citet{PM04}, the Grad-Shafranov equation is solved subject to Dirichlet and Neumann conditions at \emph{fixed} boundaries.
Mathematically, one can formulate a well-posed boundary-value problem for the Grad-Shafranov equation in the presence of a \emph{free} boundary, e.g. the sinking base of a mountain; in practice, however, this is an extremely difficult problem to solve.}

In \citet{PM07} and \citet{VM08}, Grad-Shafranov equilibria are loaded into ZEUS, a multi-purpose, time-dependent, ideal-MHD solver \citep{ZEUS2D-I,ZEUS2D-II,ZEUSMP}, and further evolved in axisymmetric \citep{PM07} and three-dimensional geometries \citep{VM08}.
The characteristic mass $\Mc$ is used to reduce the length scales of the simulated neutron star to circumvent numerical difficulties and render the simulations computationally tractable [\citet{PM07}, section~3.3; \citet{VM08}, section~2.3 and equation~(6), and section~4.6].
Two approaches are explored to augmenting the mass of a Grad-Shafranov mountain, up to $\Ma \lesssim 5.6 \Mc$: in the first approach, additional matter is injected through the outer boundary along the polar flux tube $0 \le \psi \le \psia$ [\citet{PM07}, section~4.2]; in the second approach, the density of the mountain is uniformly increased at every point, while the magnetic field is preserved [\citet{PM07}, section~4.4].
A plausible attempt to extend this latter approach to include sinking is outlined in appendix~\ref{sec:gsmatch}; ultimately this attempt proved unsuccessful, and was abandoned.
Instead, the method presented in this section uses ZEUS-MP \citep{ZEUSMP} to build magnetic mountain equilibria from scratch; this approach was first proposed in \citet{VM09GW}.

\subsection{Outline of the method}\label{ssec:outline}

\begin{figure*}
\centering
\subfloat[]{\label{fig:diagram-sink}
\includegraphics[width=0.32\textwidth]{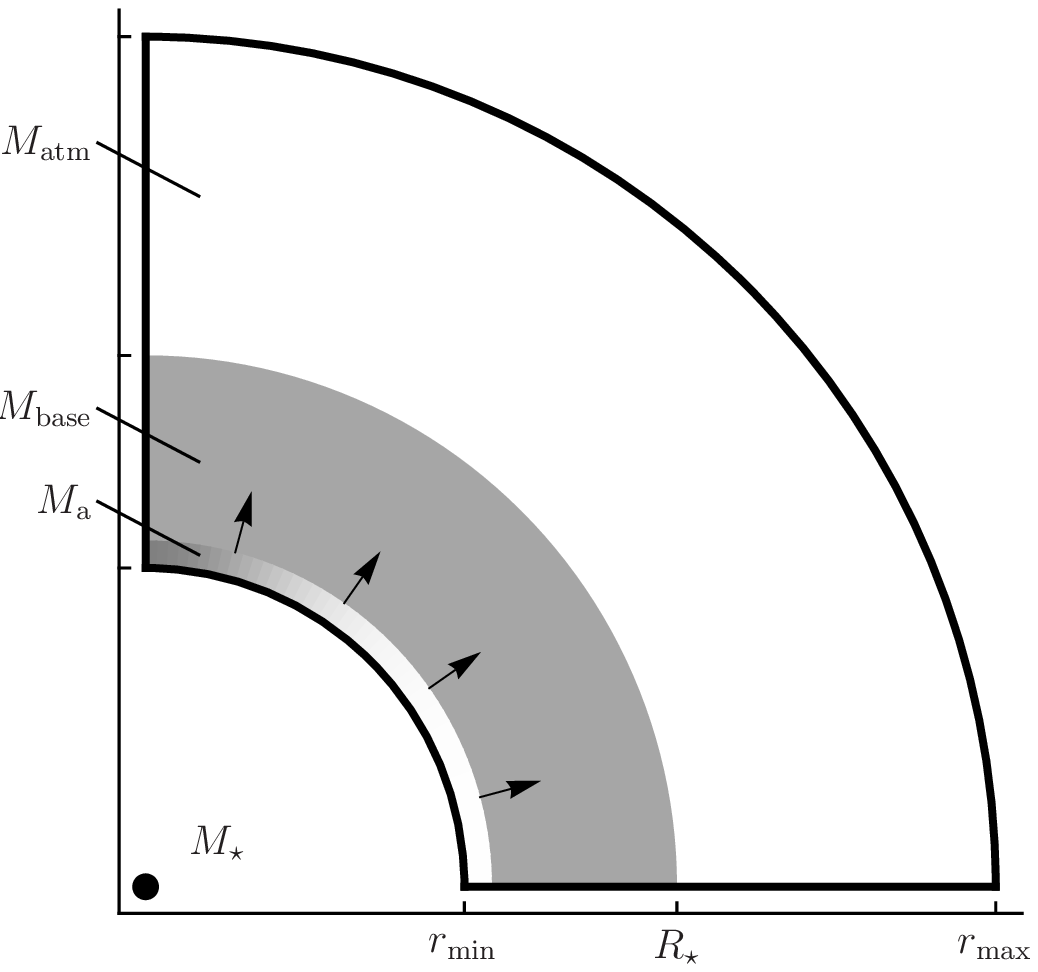}
}
\subfloat[]{\label{fig:diagram-sink-surf}
\includegraphics[width=0.32\textwidth]{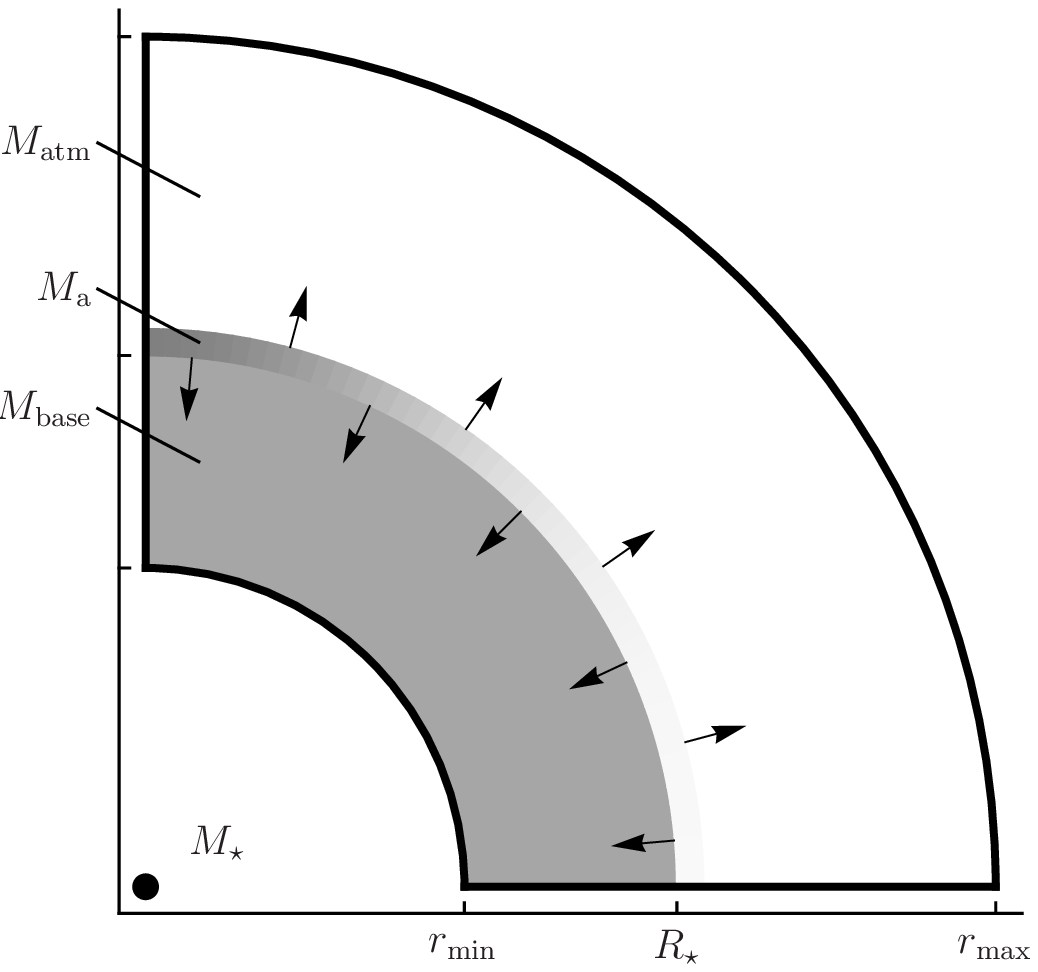}
}
\subfloat[]{\label{fig:diagram-hard}
\includegraphics[width=0.32\textwidth]{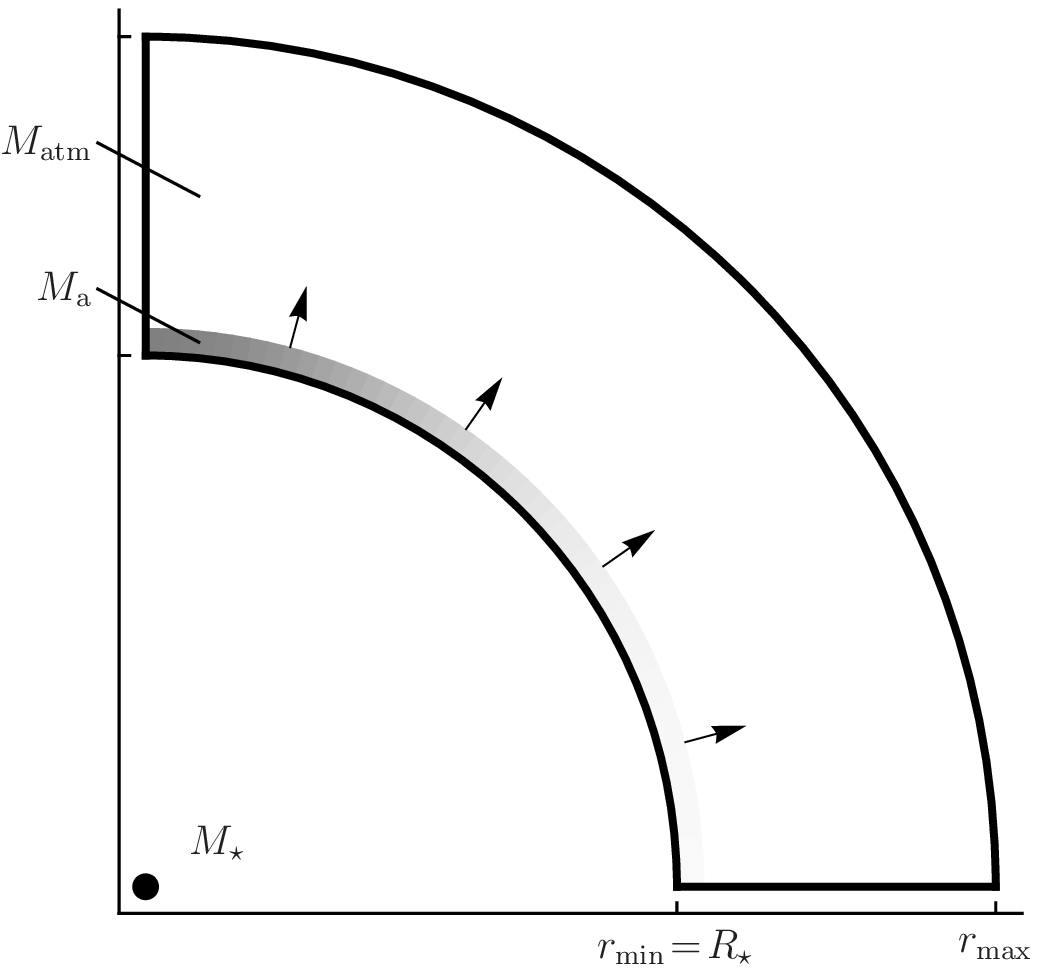}
}
\caption{\label{fig:diagram}
Diagrams illustrating schematically three mountain growth scenarios.
The simulation region is bounded by $\rmin \le r \le \rmax$, $0 \le \theta \le \pi/2$ (thick lines), and represents a quadrant of the star.
Boundary conditions assume symmetry about $\theta = 0$ and reflection at $\theta = \pi/2$.
The surface of the star is located at $r = \Rstar$.
Three sub-regions are identified.
The fluid interior beneath the surface, containing mass $\Mbase$, is shaded gray.
The region where the mountain mass $\Ma$ is injected into the simulation is shaded in a gray to white gradient; the gray is proportional to the injected flux [see equation~(\ref{eqn:massfluxrth})] as a function of $\theta$.
The outer atmosphere of the star, containing mass $\Matmos$, is unshaded.
A central gravitational point source is labelled with its mass $\Mstar$.
For mountains grown on a fluid base, $\Ma$ can be injected~\subref{fig:diagram-sink} at the inner boundary $r = \rmin$, or~\subref{fig:diagram-sink-surf} at the stellar surface $r = \Rstar$.
For mountains grown on a hard surface,~\subref{fig:diagram-hard} there is no fluid interior; the inner boundary is identical to the stellar surface $r = \rmin = \Rstar$ ($\Mbase = 0$).
See the text in section~\ref{ssec:injection}.
}
\end{figure*}

The setup of the simulations presented in this paper is described schematically in Figure~\ref{fig:diagram}.
Three numerical experiments are performed: growing onto a hard surface, growing onto a soft surface by injecting matter from below, and repeating the latter experiment by injecting matter at some altitude.

To simulate accretion, we inject matter from below, through the inner boundary of the simulation at $r = \rmin$.
One might expect a realistic simulation of accretion to add matter from above, through the outer boundary $r = \rmax$.
The two scenarios are, however, equivalent in ideal MHD; the magnetic field is frozen into the fluid, which is thus constrained to move along lines of magnetic flux.
Provided that the simulation reaches equilibrium, it becomes inconsequential, with respect to ideal MHD, through which end of a flux tube matter is added.
This is because matter cannot cross flux surfaces in ideal MHD, so the mass column $dM(\psi)$ between $\psi$ and $\psi + \delta \psi$ adjusts to reach the same hydrostatic radial profile \emph{in equilibrium}, whether it enters slowly from below or falls slowly from above.
In the presence of gravity, which (in the case of a sinking mountain) induces steep density gradients in the fluid base, the results to be presented in section~\ref{ssec:soft-vs-hard} confirm that this situation remains true; two different injection scenarios (described below) give ellipticities consistent to within 10\%.
There remains, however, the subtle and difficult question of irreversible magnetic reconnection at the grid corners, which remains unresolved (see the discussion in section~\ref{ssec:linetying}).

In practice, it is advantageous to add matter through the inner boundary, because we wish to inject along particular flux tubes, and this is easiest to do at $r = \rmin$, where the magnetic footprints are fixed in place (unlike at $r = \rmax$).
This constraint, known as magnetic line tying, contributes to the stability of the mountain \citep{Goedbloed-Poedts-2004,VM08}.
It is well justified physically, provided that $\rmin$ lies deep enough within the star, so that the fluid base (and frozen-in magnetic flux) remains relatively stationary, and is not significantly perturbed by the spreading and sinking of the mountain.
This is the case if the mass $\Mbase$ of the fluid base, initially in the region $\rmin < r < \Rstar$, is much greater than $\Ma$.
To confirm that the mountain does not greatly push the crustal material, we first calculate the fraction of $\Mbase$ contained in each grid cell, and then determine the change in this quantity between the initial and final times of the simulation; this gives the change in the spatial distribution of $\Mbase$ over the simulation, as a function of the grid cell.
For all simulations with sinking, the median change in $\Mbase$, over all grid cells, is on average $\sim 10\%$; thus, the distribution of the fluid base does not change much during accretion.
Recent molecular dynamics simulations of crystalline neutron matter, which predict a high breaking strain $\sim 0.1$ \citep{Horowitz-Kadau-2009}, also support the line-tying hypothesis.

When a mountain is grown onto a fluid base $\Mbase$, a difficulty arises.
ZEUS-MP models a single fluid, with a unique velocity field \citep{ZEUSMP}; there is no facility for simulating the movement of one fluid, the injected mountain, with respect to another fluid, the stationary crust.\footnote{
ZEUS-MP can track the concentrations of comoving components within the same fluid; we exploit this in section~\ref{ssec:injection}.}
We are left with two alternatives: to assign the same velocity to the injected mountain and the crust (the behaviour of ZEUS-MP's ``inflow'' boundary condition), or to assign a negligible or zero velocity to the injected mountain, in order to keep the base stationary.
In the first case, ZEUS-MP fails catastrophically for desirable values of the injection velocity ($\gtrsim 5\%$ of the escape velocity).
In the second case, which we study in section~\ref{sec:compare}, mountains remain subterranean and never rise to the stellar surface $r = \Rstar$.
As a check, therefore, we examine two scenarios: injection at $r = \rmin$ and $r = \Rstar$.
We show in section~\ref{ssec:soft-vs-hard} that the results in both scenarios are quantitatively alike, confirming their robustness.

Throughout this paper, we adopt the viewpoint that the accreted matter and the mountain are one and the same; the accreted mass and the mass of the mountain are identical and are both denoted by $\Ma$.
This is a matter of terminology, not physics.
There is no ``hard edge'' to the mountain; matter is accreted on all flux surfaces $0 \le \psi \le \psistar$ [see equation~(\ref{eqn:massfluxrth}) in section~\ref{ssec:injection}], not just on the polar cap $0 \le \psi \le \psia$, which contains $\sim 63\%$ of $\Ma$.
Under the assumption of ideal MHD, matter cannot spread across flux surfaces, i.e. there is no Ohmic diffusion.
We also do not model the accreted matter once it has sunk beyond the crust, as do e.g \citet{Choudhuri-Konar-2002}; see the discussion in section~\ref{sec:discuss}.

\subsection{Initial setup}\label{ssec:setup}

The initial setup of our simulations closely follows \citet{PM07} and \citet{VM08}.
The setup of ZEUS-MP\footnote{
Version 2.1.2, available from \url{http://lca.ucsd.edu/portal/codes/zeusmp2}.}
is through a set of parameters which control: the geometry of the problem, the physical phenomena to be modelled (e.g. MHD, gravity), the simulation grid and its boundary conditions, the equation of state, and the choice of timestep.
Appropriate values for these parameters are given in \citet{PM07}, section~3 and appendix~A1, and in \citet{VM08}, sections~2.2--2.3 and appendix~A.

To avoid numerical difficulties with steep magnetic field gradients, we simulate a scaled-down neutron star, where the mass $\Mstar$ and radius $\Rstar$ are artificially reduced, while the hydrostatic scale height $h_0 = \cs^2 \Rstar^2 / G \Mstar$ is kept constant \citep{PM07}.
The down-scaling transformation preserves the equilibrium shape of the mountain exactly in the small-$\Ma$ limit \citep{PM04,PM07} and has been validated approximately for $\Ma \lesssim 20 \Mc$ \citep{VM08}.
We use dimensionless units within ZEUS-MP, setting the isothermal sound $\cs$ and gravitational constant $G$ to unity, and adopting $h_0$ as the unit of length.
Table~\ref{tbl:conversions} explains how to convert between an astrophysical neutron star, the scaled-down model, and dimensionless ZEUS-MP units.

\begin{table*}
\centering
\caption{\label{tbl:conversions}
Conversion of physical quantities into dimensionless variables in the simulations.
Physical quantities are first converted to their values in the scaled-down model by multiplying by $[a(\text{Simulation})/a(\text{Astrophysical neutron star})]^n$, where $a = \Rstar / h_0$ parameterises the curvature down-scaling, and $n$ is listed in column 4.
Scaled-down physical quantities are then reexpressed in the dimensionless units of ZEUS-MP according to column 6.
The table is divided into three horizontal parts containing: stellar parameters \citep{PM04}, simulation control parameters (see Figure~\ref{fig:diagram}), and simulation outputs.
}
\begin{tabular}{@{}llllllll}
\hline

Quantity &
Symbol &
\parbox{1.6cm}{Astrophysical \\ neutron star} &
\parbox{1.8cm}{Down-scaling \\ index $n$} &
Simulation &
ZEUS-MP dimensionless unit \\

\hline

scaling ratio &
$a$ &
$1.9 \times 10^{4}$ &
none &
50 &
none \\

\hline

stellar mass &
$\Mstar$ &
$1.4 \Msun$ &
2 &
$10^{-5} \Msun$ &
$M_0 = \cs^2 h_0 / G = 8.1 \times 10^{24}~\mathrm{g}$ \\

stellar radius &
$\Rstar$ &
$10^6~\mathrm{cm}$ &
1 &
$2.7 \times 10^3~\mathrm{cm}$ &
$h_0 = \cs^2 \Rstar^2 / G \Mstar = 54~\mathrm{cm}$ \\

stellar magnetic field &
$\Bstar$ &
$10^{12}~\mathrm{G}$ &
none & &
$B_0 = \cs^2 / G^{1/2} h_0 = 7.2 \times 10^{17}~\mathrm{G}$ \\

isothermal sound speed &
$\cs$ &
$10^{8}~\mathrm{cm~s}^{-1}$ &
none & &
$\cs$ \\

critical mass &
$\Mc$ &
$1.2 \times 10^{-4} \Msun$ &
4 &
$6.1 \times 10^{-15} \Msun$ &
$M_0$ \\

\hline

inner boundary &
$\rmin$ & &
1 &
see Table~\ref{tbl:problems} &
$h_0$ \\

outer boundary &
$\rmax$ & &
1 &
$1.2 \Rstar$ &
$h_0$ \\

accreted mass &
$\Ma$ & &
4 &
see Table~\ref{tbl:problems} &
$M_0$ \\

mass of outer atmosphere &
$\Matmos$ & &
4 &
$5 \times 10^{-6} \Msun$ &
$M_0$ \\

mass of fluid base &
$\Mbase$ & &
4 &
$10 \Ma$ &
$M_0$ \\

\hline

mountain density &
$\rho \Xmnt$ & &
1 & &
$\rho_0 = M_0 / h_0^3 = 5.2 \times 10^{19}~\mathrm{g~cm}^{-3}$ \\

magnetic field &
$\Bb$ & &
none & &
$B_0$ \\

ellipticity &
$\epsilon$ & &
2 & &
none \\

time &
$t$ & &
none & &
$t_0 = h_0 / \cs = 5.4 \times 10^{-7}~\mathrm{s}$ \\

\hline
\end{tabular}
\end{table*}

The simulations are performed on an axisymmetric rectangular grid with $N_{r}$ cells spaced logarithmically in $r$, and $N_{\theta} = 64$ cells spaced linearly in $\theta$.
The logarithmic spacing in $r$ is determined by the ratio $\Delta r_{N_r - 1} / \Delta r_0$ of the maximum to minimum radial grid spacing (see appendix~\ref{sec:logspacing}).
This ratio is chosen large enough to concentrate grid resolution near the inner boundary, but small enough to ensure reasonable run times.
We set $\rmax = 1.2 \Rstar = 60 h_0$ to give the mountain ample room to expand without meeting the outer boundary, and stipulate reflecting boundary conditions at $\theta = 0$ and $\theta = \pi/2$, ``inflow'' boundary conditions at $r = \rmin$, and ``outflow'' boundary conditions at $r = \rmax$; more details can be found in \citet{PM07}.
The magnetic field is initially that of a dipole, and $\Bstar$ is its magnitude at the polar surface.

A gravitational point source $\Mstar$ is placed at $r = 0$, and self-gravity is ignored.
The density field is initialised to be the static atmosphere of an isothermal fluid with no self-gravity:
\begin{equation}\label{eqn:initrho}
\rho(t=0,r) = \rhosurf \exp \left[ \frac{G \Mstar}{\cs^2}
\left( \frac{1}{r} - \frac{1}{\Rstar} \right) \right] \,.
\end{equation}
Ideally the region $r > \Rstar$ should start evacuated, but ZEUS-MP requires the density to be nonzero everywhere, so we set $\Matmos = 5 \times 10^{-6} \Msun$ (approximately $4\%$ of the mass of the smallest mountain; see Table~\ref{tbl:problems}).
Integrating equation~(\ref{eqn:initrho}) over the region $r > \Rstar$ (see Figure~\ref{fig:diagram}) fixes the density at the stellar surface $\rho_{\Rstar}$ in terms of $\Matmos$.
In contrast, we require the mass of the fluid base $\Mbase$ (when a soft surface is being modelled) to be much larger than the mass of the mountain, as discussed in section~\ref{ssec:outline}.
In all runs, we choose $\Mbase/\Ma \approx 10$. Integrating equation~(\ref{eqn:initrho}) over the region containing $\Mbase$ then fixes $\rmin$.

\subsection{Injection procedure}\label{ssec:injection}

ZEUS-MP's ``inflow'' boundary condition permits injection at the edge of the simulation volume.
To enable injection at $r = \Rstar$, as in Figure~\subref*{fig:diagram-sink-surf}, we implement a more flexible custom procedure, and use the built-in ``inflow'' condition only to tie the magnetic flux at $r = \rmin$.
We describe the procedure briefly below; further details are in appendix~\ref{sec:extrainj}.

\begin{figure}
\centering
\includegraphics[width=\columnwidth]{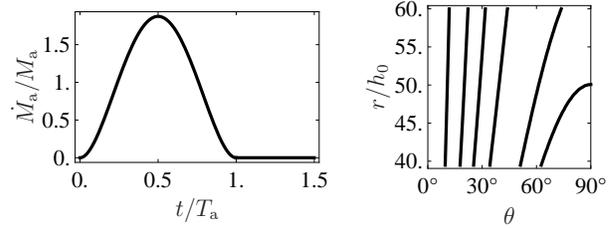}
\caption{\label{fig:massflux-t-rth}
(left) The accretion rate $\Madot(t)$, given by equation~(\ref{eqn:massfluxt}).
(right) The initial mass-flux distribution $\partial M / \partial \psi$, given by equation~(\ref{eqn:massfluxrth}).
Contours are at (right to left) 0.05, 0.1, 0.3, 0.5, 0.7, and 0.9 of the maximum.
}
\end{figure}

We inject mass $\Ma$ into an injection region $\rinj < r < \rinj + \drinj$, $0 < \theta < \pi/2$, over a time interval $0 < t < \Tinj$.
(We set $\drinj = 0.1 h_0$ throughout.)
The flux of accreted matter at time $t$ entering a point $(r,\theta)$ in the injection region is given by
\begin{equation}\label{eqn:massflux}
\frac{ \partial^3 \Ma }{ \partial t \partial r \partial \theta }(t,r,\theta) \propto
\Madot(t) \frac{ \partial \Ma }{ \partial \psi }(r,\theta) \,.
\end{equation}
where we choose
\begin{equation}\label{eqn:massfluxt}
\Madot(t) \propto t^2 (\Tinj - t)^2 \,,
\end{equation}
and
\begin{equation}\label{eqn:massfluxrth}
\frac{ \partial \Ma }{ \partial \psi }(r,\theta) \propto
\exp ( -b \Rstar r^{-1} \sin^2\theta ) \,.
\end{equation}
The normalisation of equation~(\ref{eqn:massflux}) is chosen so that, for each simulation, the mass of the mountain is equal to $\Ma$ at time $t = \Tinj$, i.e. $\Ma(t = \Tinj) \equiv \Ma$.
After time $t = \Tinj$, no further mass is added, but we evolve the system up to $t = \tmax = 1.5\Tinj$ to test the stability of the mountain obtained.

Equation~(\ref{eqn:massfluxt}) determines the rate of accretion; it is plotted in Figure~\ref{fig:massflux-t-rth}~(left).
The functional form was chosen to ensure numerical stability in ZEUS-MP, and has no particular astrophysical justification, except to ensure that a mountain builds up to its target mass smoothly over the time scale $\Tinj$.
For this reason, it is a smooth bell-shaped function, designed to avoid any discontinuity in the accretion rate, which might excite undesired oscillations in the fluid or provoke numerical instabilities.

Equation~(\ref{eqn:massfluxrth}) gives a mass-flux distribution consistent with that of \citet{PM04}; it is plotted in Figure~\ref{fig:massflux-t-rth}~(right).
It does not attempt to model the interaction of the accreted matter with the magnetosphere, from which the mass-flux distribution would originate; instead, it is chosen such that the majority ($\sim 63\%$) of the accreted matter falls on the polar cap $0 \le \psi \le \psia$.
The parameter $b = \psistar / \psia = 3$ determines the polar cap radius $\Rstar \sin^{-1} (b^{-1/2})$.
It is determined astrophysically by disk-magnetosphere force balance, and is related to the stellar magnetic field via $b \propto \psistar \propto \Bstar$ \citep{PM04}.
In this theoretical paper, however, we treat $b$ (and therefore $\Bstar$) as a free parameter, and do not attempt a self-consistent solution of the disk-magnetosphere system \citep[see, e.g.,][]{Romanova-etal-2008}.
With this freedom, $b$ is chosen unrealistically large to preserve numerical stability \citep{PM04}.

We use ZEUS-MP's multi-species tracking facility \citep{ZEUSMP} to record, throughout the simulation, the fraction of the density, $0 \le \Xmnt(t,r,\theta) \le 1$, that originates from accretion (i.e. added at $t > 0$ via the injection procedure), as opposed to from the initial configuration at $t = 0$.
This allows us to track the spread of the mountain independently of the motion of the remaining (displaced) stellar matter.

We require that the mountain grows quasistatically, in the sense that the accretion timescale $\Tinj$ is always much greater than $\talfv$, the characteristic pole-equator crossing time of an Alfv\'en wave.
Following \citet{VM08}, we compute the crossing time at $t = 0$, $r = \Rstar$: from the Alfv\'en speed $\valfv = \Bstar / (4\pi \rhosurf)^{1/2} \approx 0.2 \cs$ (see Table 1), we obtain $\talfv = \pi \Rstar / (2 \valfv) \approx 400 t_0$.
The condition $\talfv \ll \Tinj$ is verified by comparison with the values for $\Tinj$ listed in Table~\ref{tbl:problems}.
The condition also implies that the magnetostatic limit always holds: the ratio $\Bstar \Ma / \Madot \gg \pi^{3/2} \Rstar \rhosurf^{1/2} \approx 2 \times 10^{8}~\mathrm{G}~\mathrm{s}$, and from Tables 1 and 2, $\Bstar \Ma / \Madot \approx \Bstar \Tinj \gtrsim 3 \times 10^{9}~\mathrm{G}~\mathrm{s}$.

For mountains grown on a hard surface, we additionally set the velocity $\vb(t,r,\theta)$ within the injection region, such that the accreted matter is always given a fixed speed $\vinj = 10^{-4} \cs$ parallel to the magnetic field $\Bb(t,r,\theta)$.
The value of $\vinj$ should be a small fraction of the escape speed $v_\mathrm{esc} \approx 4.1 \cs$, so there is negligible mass lost through the outer boundary (see section~\ref{ssec:probhiii}).
We find that setting $\vb$ carefully is critical to stability.

\section{Comparing mountains grown on hard and soft bases}\label{sec:compare}

\begin{table*}
\centering
\caption{\label{tbl:problems}
Simulations of magnetic mountains presented in this paper.
The accompanying parameters are:
the target accreted mass $\Ma$, in units of $\Mc$ and $\Msun$;
the number of grid cells in the $r$ direction $N_{r}$;
the radius of the inner radial boundary $\rmin$;
the ratio of the maximum to minimum radial grid spacing $\Delta r_{N_r - 1} / \Delta r_0$;
the injection radius $\rinj$;
the injection velocity $\vinj$;
the injection time $\Tinj$; and
the total (successfully completed) simulation time $\tmax$.
}
\begin{tabular}{@{}lllllllllll}
\hline
Simulation &
$\frac{\Ma}{\Mc}$ &
$\frac{\Ma}{\Msun}$ &
$N_{r}$ &
$\frac{\rmin}{h_0}$ &
$\frac{ \Delta r_{N_r - 1} }{ \Delta r_0 }$ &
$\rinj$ &
$\frac{\vinj}{\cs}$ &
$\frac{\Tinj}{t_0}$ &
$\frac{\tmax}{\Tinj}$
\\
\hline
$\probhi$ & $1$ & $1.2 \times 10^{-4}$ & $64$ & $50.0$ & $200$ & $\rmin = \Rstar$ & $10^{-4}$ & $5 \times 10^{3}$ & $1.5$ \\
$\probhii$ & $10$ & $1.2 \times 10^{-3}$ & $64$ & $50.0$ & $200$ & $\rmin = \Rstar$ & $10^{-4}$ & $5 \times 10^{3}$ & $1.5$ \\
$\probhiii$ & $100$ & $1.2 \times 10^{-2}$ & $64$ & $50.0$ & $200$ & $\rmin = \Rstar$ & $10^{-4}$ & $5 \times 10^{3}$ & $1.5$ \\
$\probhiiii$ & $1000$ & $1.2 \times 10^{-1}$ & $64$ & $50.0$ & $200$ & $\rmin = \Rstar$ & $10^{-4}$ & $5 \times 10^{3}$ & $1.5$ \\
\hline
$\probsi$ & $1$ & $1.2 \times 10^{-4}$ & $96$ & $44.7$ & $306$ & $\rmin$ & $0$ & $5 \times 10^{3}$ & $1.5$ \\
$\probsii$ & $10$ & $1.2 \times 10^{-3}$ & $112$ & $42.8$ & $344$ & $\rmin$ & $0$ & $5 \times 10^{3}$ & $1.5$ \\
$\probsiii$ & $100$ & $1.2 \times 10^{-2}$ & $120$ & $41.1$ & $378$ & $\rmin$ & $0$ & $5 \times 10^{3}$ & $1.5$ \\
$\probsiiii$ & $1000$ & $1.2 \times 10^{-1}$ & $128$ & $39.5$ & $410$ & $\rmin$ & $0$ & $5 \times 10^{3}$ & $1.5$ \\
\hline
$\probssi$ & $1$ & $1.2 \times 10^{-4}$ & $96$ & $44.7$ & $306$ & $\Rstar$ & $0$ & $5 \times 10^{3}$ & $1.5$ \\
$\probssii$ & $10$ & $1.2 \times 10^{-3}$ & $112$ & $42.8$ & $344$ & $\Rstar$ & $0$ & $1 \times 10^{4}$ & $1.5$ \\
$\probssiii$ & $100$ & $1.2 \times 10^{-2}$ & $120$ & $41.1$ & $378$ & $\Rstar$ & $0$ & $2 \times 10^{4}$ & $1.5$ \\
$\probssiiii$ & $1000$ & $1.2 \times 10^{-1}$ & $128$ & $39.5$ & $410$ & $\Rstar$ & $0$ & $8 \times 10^{4}$ & $0.35$ \\
\hline
\end{tabular}
\end{table*}

\begin{figure}
\centering
\includegraphics[width=\columnwidth]{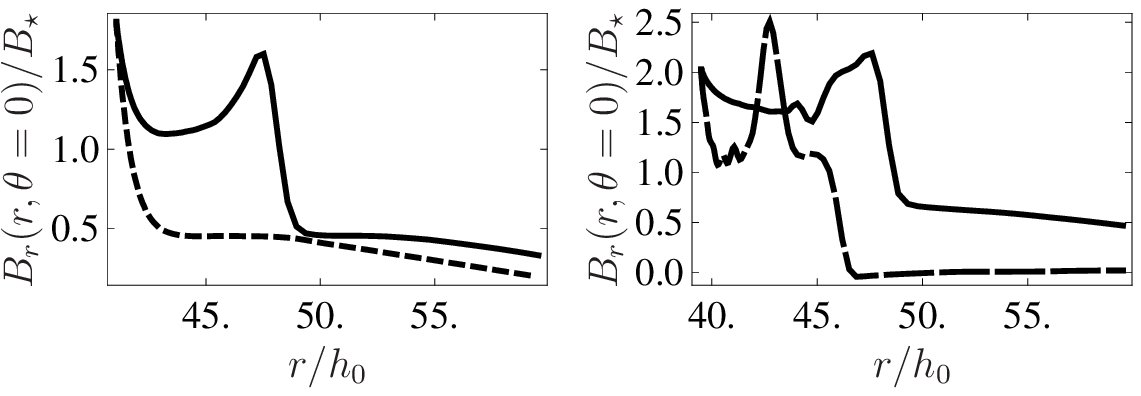}
\caption{\label{fig:sink_surf_instab}
Radial component of the magnetic field $B_r$ along $\theta = 0$: for simulation $\probssiii$ (left), at $t = 0.45\Tinj$ (solid) and $t = 0.85\Tinj$ (dotted); and for simulation $\probssiiii$ (right), at $t = 0.24\Tinj$ (solid), and its time of failure $t = 0.35\Tinj$ (dashed).
}
\end{figure}

Table~\ref{tbl:problems} lists the parameters of the simulations presented in this paper.
Mountains grown on a hard surface are labelled $\mathcal{H}(\Ma/\Mc)$.
Mountains grown on a fluid base are labelled $\mathcal{S}(\rinj,\Ma/\Mc)$, where the injection radius $\rinj$ may be either $\rmin$ or $\Rstar$.
The parameters of each run are chosen to grow a mountain with a particular target mass, $\Ma$.
We choose four values for $\Ma$ in the range $10^{-4}$ -- $10^{-1}\Msun$.
These values are chosen to demonstrate the ability of the injection procedure to generate stable mountains over a wide range of masses.
This range also encompasses the range of $\Ma$ of real accreting neutron stars (see Table~\ref{tbl:parameters}).
The main source of uncertainty is the accretion efficiency \citep{vdHeuvel-Bitzaraki-1995}, which may be as low as $\sim 5\%$ \citep{Tauris-etal-2000}; this is reflected in the chosen range of $\Ma$.

The CPU time required for each run was, on average, $\sim 10^{-2}$ seconds per grid cell per unit $t_0$ of simulation time.
For $\probss$-type simulations, one must scale $\Tinj$ with $\Ma$ to prevent numerical instabilities.
Even so, run $\probssiiii$ does not complete; ZEUS-MP aborts at $t \approx 0.35 \Tinj$, when the adaptive time-step shrinks below its allowed minimum.
Figure~\ref{fig:sink_surf_instab} shows that this behaviour arises when $B_r$ diverges at $r \lesssim \rinj$ along the boundary $\theta = 0$: for $\Ma = \Miii\Mc$, $B_r$ threatens to break out for $t \lesssim 0.5 \Tinj$ but ultimately settles down to the equilibrium configuration before $t = \Tinj$, whereas for $\Ma = \Miiii\Mc$, it grows uncontrollably up to the time of failure.

\subsection{Verification}\label{ssec:verification}

\begin{figure}
\centering
\subfloat[$\probh$]{\label{fig:mnt-mass-hard}
\includegraphics[width=0.49\columnwidth]{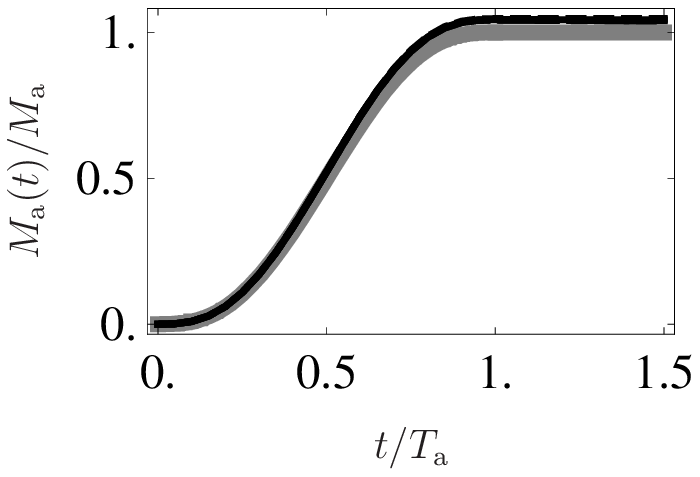}
}
\subfloat[$\probh$]{\label{fig:star-mass-hard}
\includegraphics[width=0.49\columnwidth]{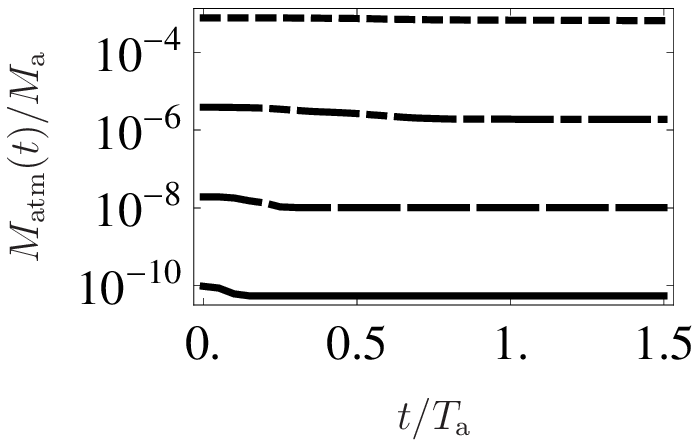}
}
\\
\subfloat[$\probs$]{\label{fig:mnt-mass-sink}
\includegraphics[width=0.49\columnwidth]{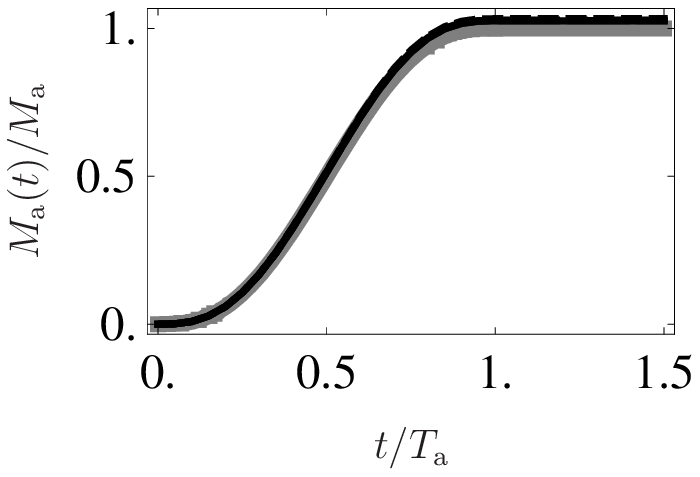}
}
\subfloat[$\probs$]{\label{fig:star-mass-sink}
\includegraphics[width=0.49\columnwidth]{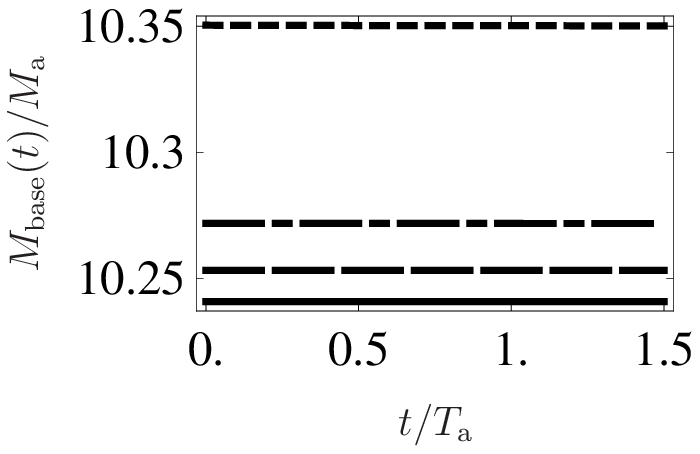}
}
\\
\subfloat[$\probss$]{\label{fig:mnt-mass-sink-surf}
\includegraphics[width=0.49\columnwidth]{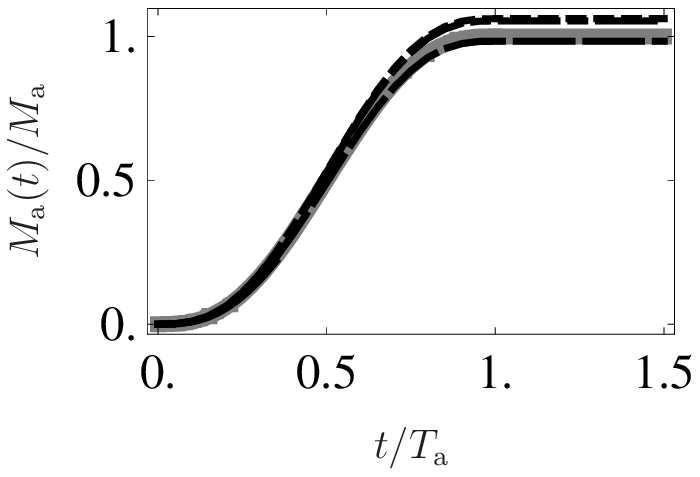}
}
\subfloat[$\probss$]{\label{fig:star-mass-sink-surf}
\includegraphics[width=0.49\columnwidth]{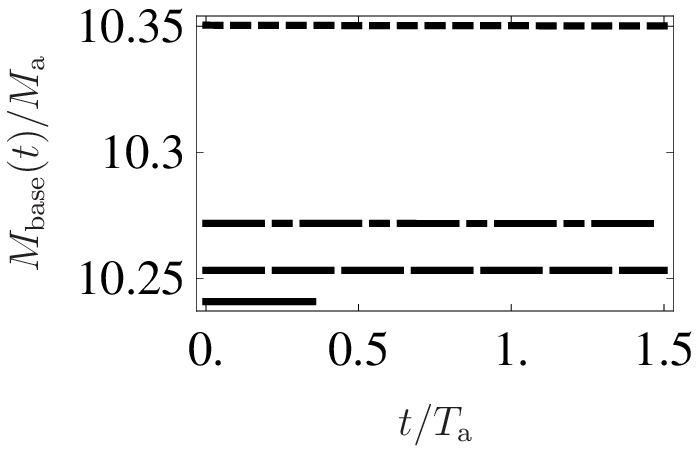}
}
\caption{\label{fig:mass}
Accreted mass $\Ma(t)$ and masses in the stellar atmosphere $\Matmos(t)$ and fluid base $\Mbase(t)$, plotted in black, for $\Ma/\Mc = \Mi$ (dotted), $\Mii$ (dot-dashed), $\Miii$ (dashed), and $\Miiii$ (solid).
The injected mass from equation~(\ref{eqn:massfluxt}) is over-plotted in gray.
The labels beneath each panel indicate a hard-surface ($\probh$) or soft-surface [$\probs$ or $\probss$] run; see Table~\ref{tbl:problems}.
The short solid line at the bottom of Figure~\subref{fig:star-mass-sink-surf} is from the aborted run $\probssiiii$
}
\end{figure}

We first check that, for each mountain, (i) we accumulate the correct total mass $\Ma$, with minimal loss through the outer boundary; (ii) the mass above the surface, $\Matmos$, remains much smaller than $\Ma$; and (iii) for mountains with sinking, the mass in the fluid base, $\Mbase$, remains large compared to $\Ma$, so that the magnetic line-tying condition at $r = \rmin$ is a good approximation.
Figures~\subref*{fig:mnt-mass-hard},~\subref*{fig:mnt-mass-sink}, and~\subref*{fig:mnt-mass-sink-surf} show $\Ma(t) = \int_{V} dV \rho \Xmnt$ integrated over the simulation volume $V$ at time $t$.
We see that the mountains achieve their target mass, which remains in the simulation for $t > \Tinj$.
The injected mass $\Ma(t)/\Ma$, found by integrating equation~(\ref{eqn:massfluxt}) with respect to time, is plotted alongside in grey; the two curves overlap.
Figure~\subref*{fig:star-mass-hard} shows $\Matmos$ for the hard-surface experiment; it is always small.
Figures~\subref*{fig:star-mass-sink} and~\subref*{fig:star-mass-sink-surf} show $\Mbase$ for the soft-surface experiments; it always exceeds $\approx 10 \Ma$, as desired.
For all simulations where $\Mbase > 0$, the total fraction of $\Mbase$ lost through the outer boundary is $\ll 0.01 \%$, consistent with \citet{VM08}.

\begin{figure}
\centering
\subfloat[$\probh$]{\label{fig:energy-hard}
\includegraphics[width=0.49\columnwidth]{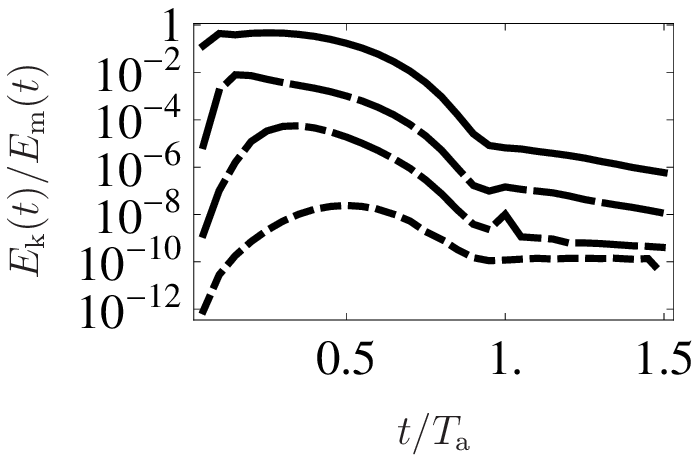}
}
\\
\subfloat[$\probs$]{\label{fig:energy-sink}
\includegraphics[width=0.49\columnwidth]{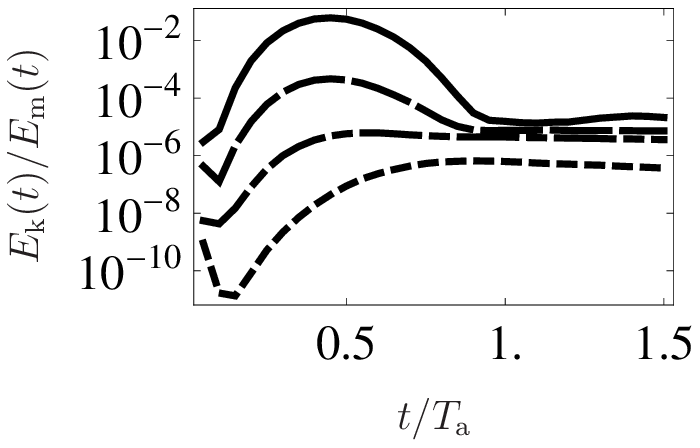}
}
\subfloat[$\probss$]{\label{fig:energy-sink-surf}
\includegraphics[width=0.49\columnwidth]{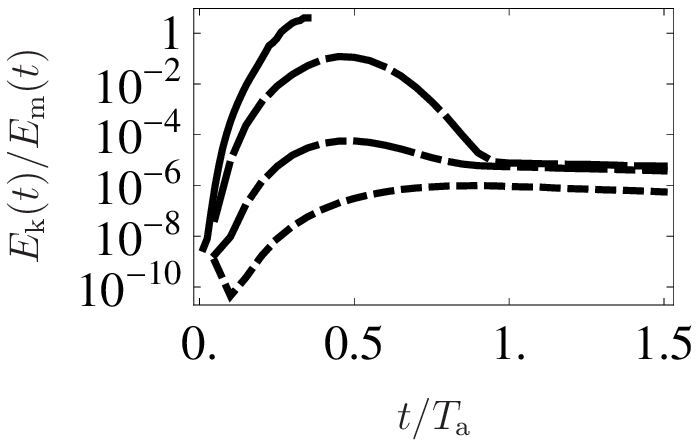}
}
\caption{\label{fig:energy}
Kinetic energy $\Ek(t)$, normalised by the magnetic energy $\Em(t)$, for $\Ma/\Mc = \Mi$ (dotted), $\Mii$ (dot-dashed), $\Miii$ (dashed), and $\Miiii$ (solid).
The labels beneath each panel indicate a hard-surface ($\probh$) or soft-surface ($\probs$) run; see Table~\ref{tbl:problems}.
}
\end{figure}

We next check that accretion takes place in the magnetostatic limit, i.e. that the total kinetic energy $\Ek = \int_{V} dV \rho |\vb|^2 / 2$ is small compared to the total magnetic energy $\Em = \int_{V} dV |\Bb|^2 / (8\pi)$.
Figure~\ref{fig:energy} shows the ratio of $\Ek$ to $\Em$ as a function of time.
We see that $\Ek/\Em$ tends to increase with $\Ma$ but typically never rises above 1\%, except in $\probhiiii$ and the incomplete run $\probsiiii$.
After accretion stops, at $t = \Tinj$, $\Ek/\Em$ typically falls to less than $10^{-4}$.

\begin{figure}
\centering
\includegraphics[width=0.23\textwidth]{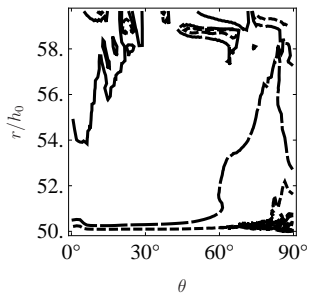}
\caption{\label{fig:divergence}
Contours of the absolute, normalised divergence of $\Bb$ of mountain $\probhiiii$ at $t = \Tinj$.
Contour levels are at $5 \times 10^{-2}$ (solid), $10^{-2}$ (dashed), $5 \times 10^{-3}$ (dotted).
}
\end{figure}

Magnetic field transport in ZEUS-MP is divergence-free by construction \citep{ZEUSMP}, but it is worth checking whether this property is preserved by the injection algorithm.
We find that the mean value of $|\nabla\cdot\Bb| / \sum_{i} (|B_{i}|/dx^{i})$ is initially $\lesssim 6 \times 10^{-3}$, and increases by a factor of 3.5 at most over the run.
Figure~\ref{fig:divergence} shows contours of the normalised $|\nabla\cdot\Bb|$ for an illustrative mountain.

\subsection{Illustrative example}\label{ssec:probhiii}

\begin{figure*}
\centering
\subfloat[$t=0.1\Tinj$]{\label{fig:dens-hard-2-t3}
\includegraphics[width=0.23\textwidth]{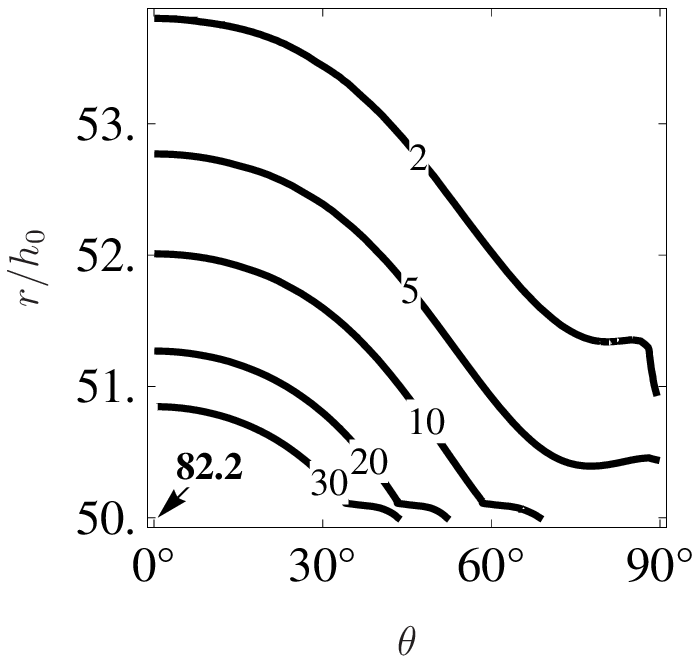}
}
\subfloat[$t=0.5\Tinj$]{\label{fig:dens-hard-2-t11}
\includegraphics[width=0.23\textwidth]{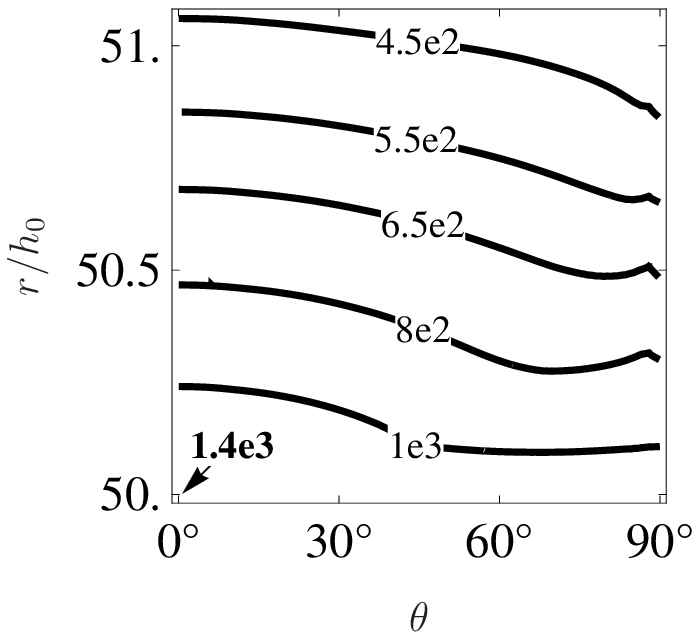}
}
\subfloat[$t=\Tinj$]{\label{fig:dens-hard-2-t21}
\includegraphics[width=0.23\textwidth]{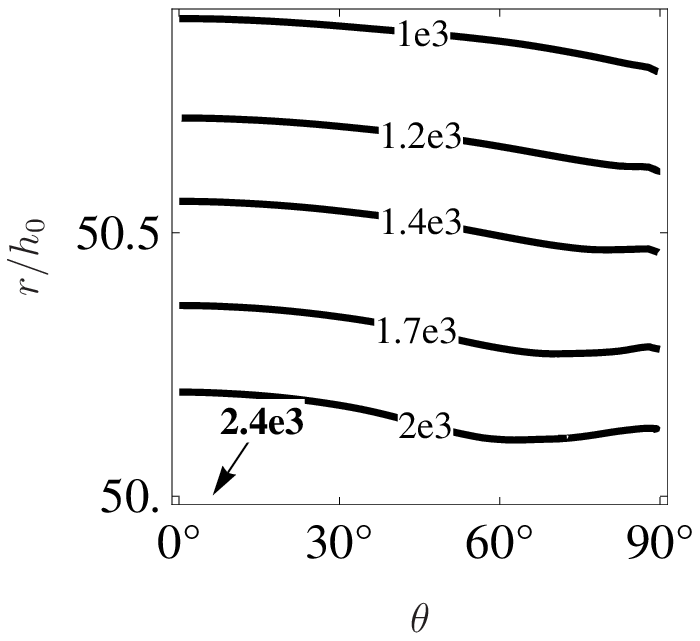}
}
\subfloat[$t=\Tinj$]{\label{fig:dens-hard-2-t31}
\includegraphics[width=0.23\textwidth]{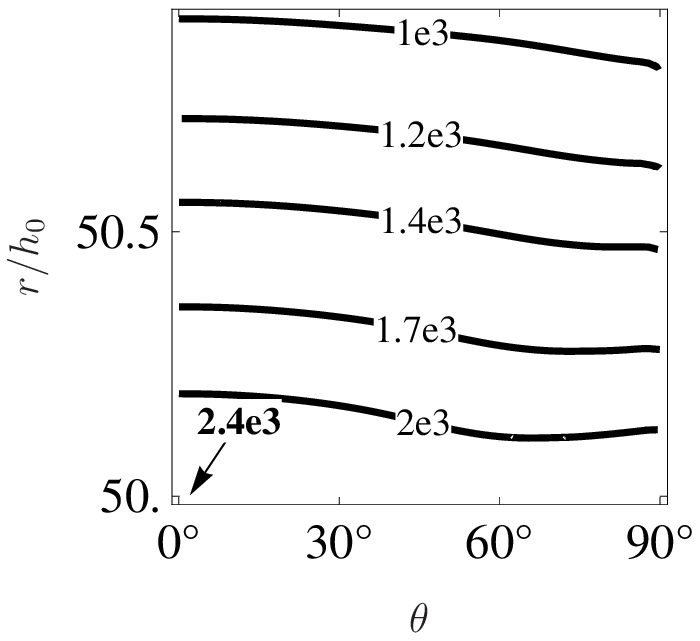}
}
\\
\subfloat[$t=0.1\Tinj$]{\label{fig:flux-hard-2-t3}
\includegraphics[width=0.23\textwidth]{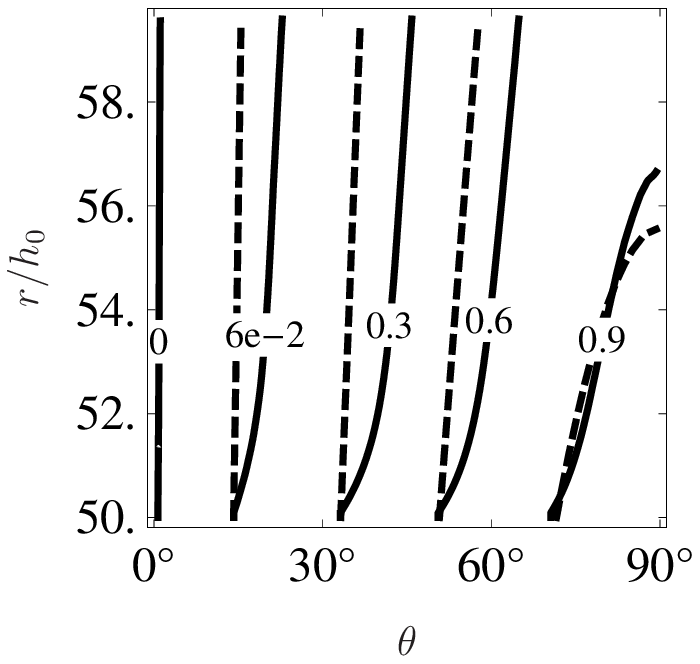}
}
\subfloat[$t=0.5\Tinj$]{\label{fig:flux-hard-2-t11}
\includegraphics[width=0.23\textwidth]{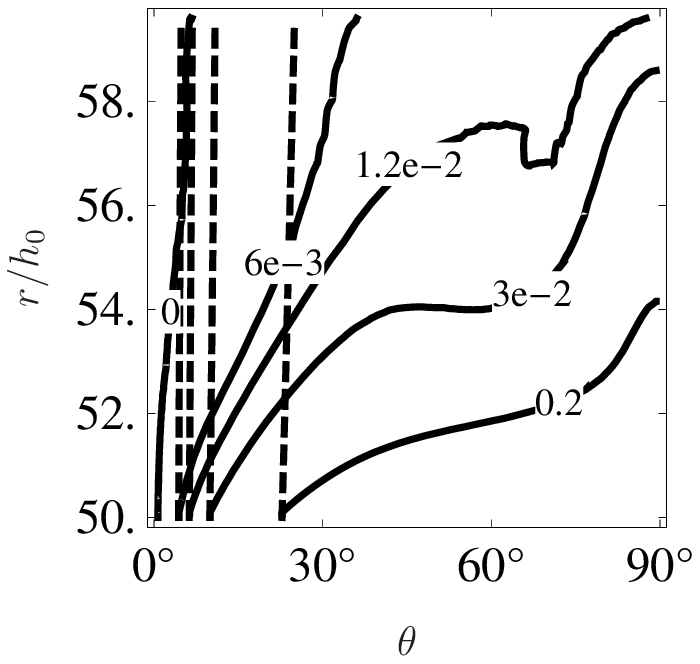}
}
\subfloat[$t=\Tinj$]{\label{fig:flux-hard-2-t21}
\includegraphics[width=0.23\textwidth]{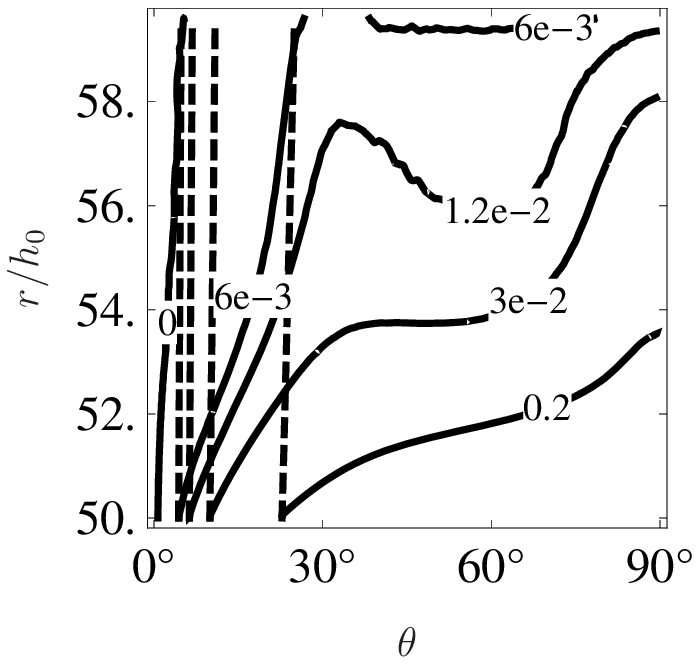}
}
\subfloat[$t=\Tinj$]{\label{fig:flux-hard-2-t31}
\includegraphics[width=0.23\textwidth]{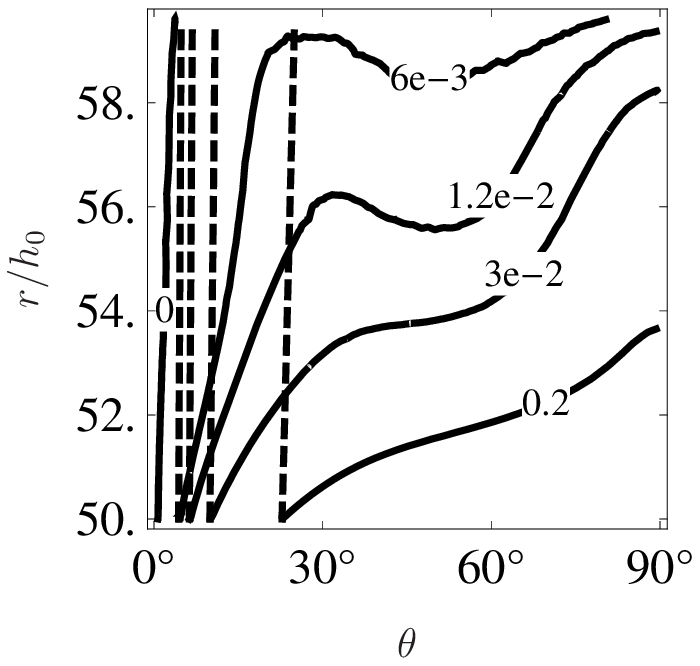}
}
\caption{\label{fig:probhiii}
Hydromagnetic structure of hard-surface mountain $\probhiii$ at times $t/\Tinj = 0.1$, 0.5, 1.0, and 1.5.
(Top row) Contours of accreted density $\rho \Xmnt / \rhosurf$; the maximum is indicated with a small arrow and labelled in bold.
(Bottom row) Contours of magnetic flux $\psi / \psistar$, at the labelled times (solid), and at $t = 0$ (dashed).
The dotted contours meet their solid equivalents on the left vertical axis.
Note that the scale of the $r$ axis differs between Figures~\subref{fig:dens-hard-2-t3}--\subref{fig:dens-hard-2-t31}.
The change in the contour scale between Figure~\subref{fig:flux-hard-2-t3} and Figures~\subref{fig:flux-hard-2-t11}--\subref{fig:flux-hard-2-t31} is due to the difference in $\psi$ between $t = 0$, where the magnetic field is dipolar, and subsequent times when the magnetic field is distorted.
}
\end{figure*}

We choose mountain $\probhiii$, grown on a hard surface with $\Ma = \Miii\Mc$, to illustrate the general evolution of a magnetic mountain during accretion.
The top row of Figure~\ref{fig:probhiii} shows contours of the mountain density $\rho \Xmnt$, normalised by the initial surface density $\rhosurf$, at four different times.
The lower row shows the magnetic flux $\psi$, normalised by $\psistar = \Bstar \Rstar^2 / 2$, at the same times.
Matter is added predominately at the pole, as determined by $dM/d\psi$.
In the early stages of accretion ($t = 0.1\Tinj$), the magnetic field is only slightly disturbed.
As accretion progresses, the mountain spreads towards the equator, dragging the frozen-in magnetic field with it.
The angular span of the $\psi$ contours is compressed from $\sim 70^\circ$ [Figure~\subref*{fig:flux-hard-2-t3}] to $\sim 20^\circ$ [Figure~\subref*{fig:flux-hard-2-t11}].
At the half-way point ($t = 0.5\Tinj$), the flux is significantly displaced from its initial configuration, but remains anchored to the inner boundary at $r = \rmin$, demonstrating magnetic line tying.
We see, in the $\psi / \psistar = 1.2 \times 10^{-2}$ contour, the early formation of the magnetic ``tutu'' configuration, observed in \citet{PM04,PM07} for $\Ma = 10^{-5} \Msun$.

At $t = \Tinj$, the mountain reaches its target mass ($\Ma = \Miii\Mc$ in Figure~\ref{fig:probhiii}).
Despite sliding towards the equator, the accreted matter still exhibits a noticeable variation in density with respect to $\theta$; a polar mountain is formed.
The tutu configuration of the magnetic field is clearly visible; see for comparison Figures~2 and~4(a) of \citet{PM04}.
This equilibrium state remains largely unchanged when we run the simulation for an additional $0.5\Tinj$, during which no further mass is added.

\subsection{Sinking}\label{ssec:soft-vs-hard}

\begin{figure*}
\centering
\subfloat[$\probhii$]{\label{fig:dens-hard-1-t21}
\includegraphics[width=0.23\textwidth]{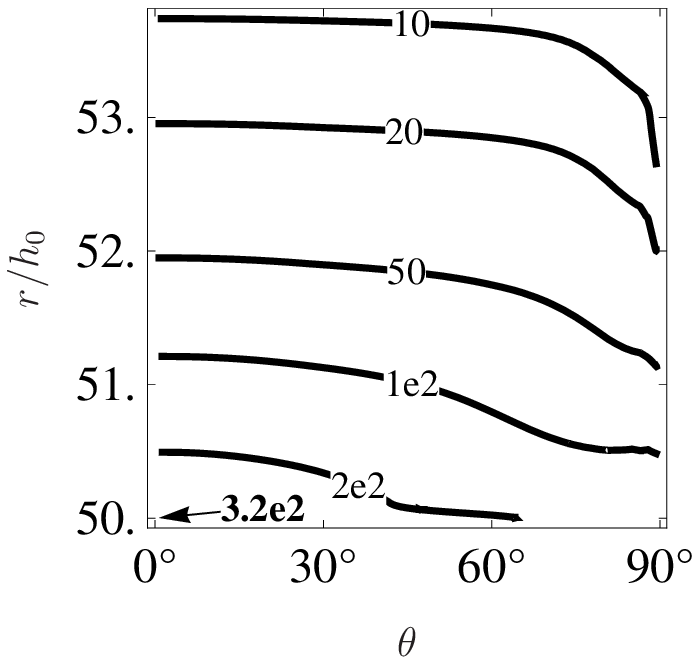}
}
\subfloat[$\probsii$]{\label{fig:dens-sink-1-t21}
\includegraphics[width=0.23\textwidth]{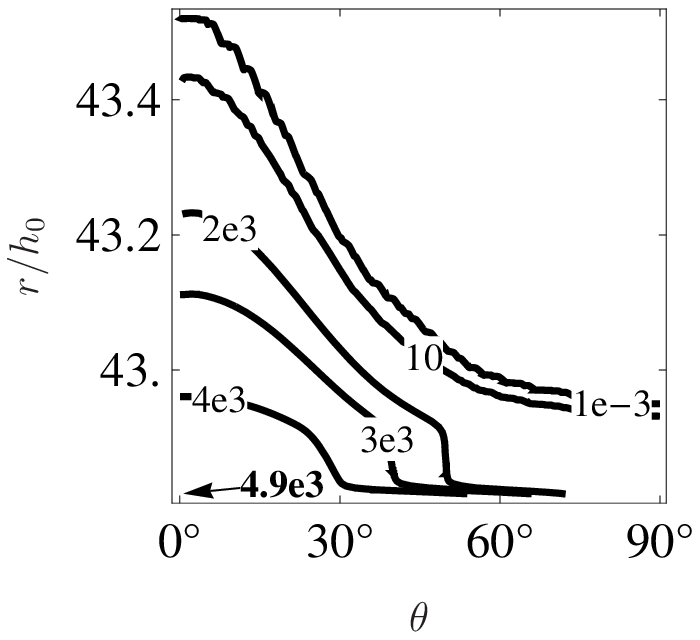}
}
\subfloat[$\probssii$]{\label{fig:dens-sink-surf-1-t21}
\includegraphics[width=0.23\textwidth]{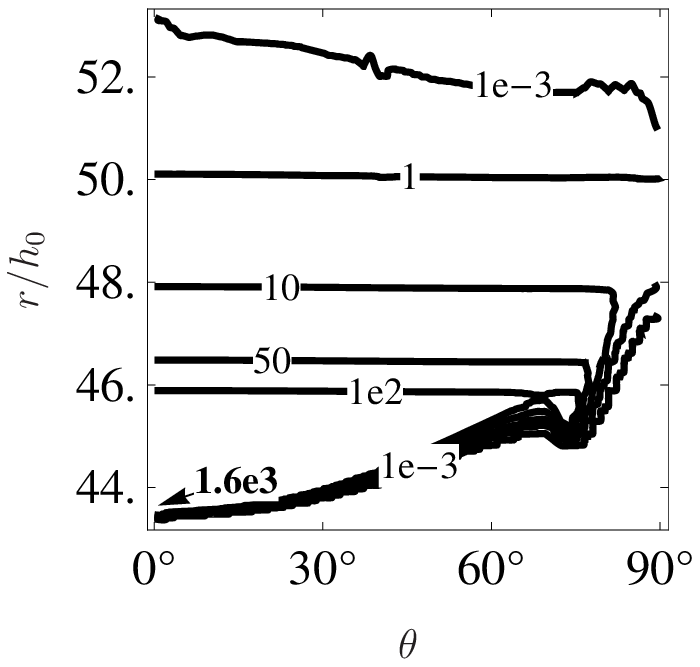}
}
\\
\subfloat[$\probhii$]{\label{fig:flux-hard-1-t21}
\includegraphics[width=0.23\textwidth]{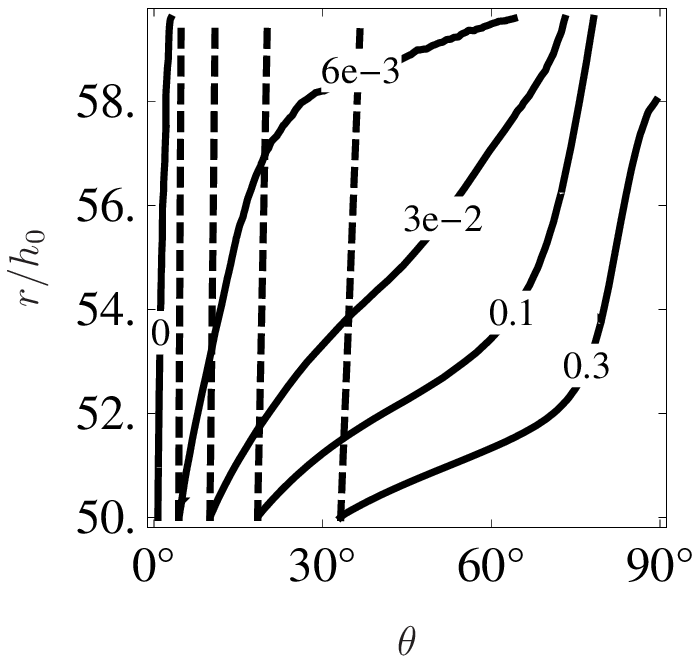}
}
\subfloat[$\probsii$]{\label{fig:flux-sink-1-t21}
\includegraphics[width=0.23\textwidth]{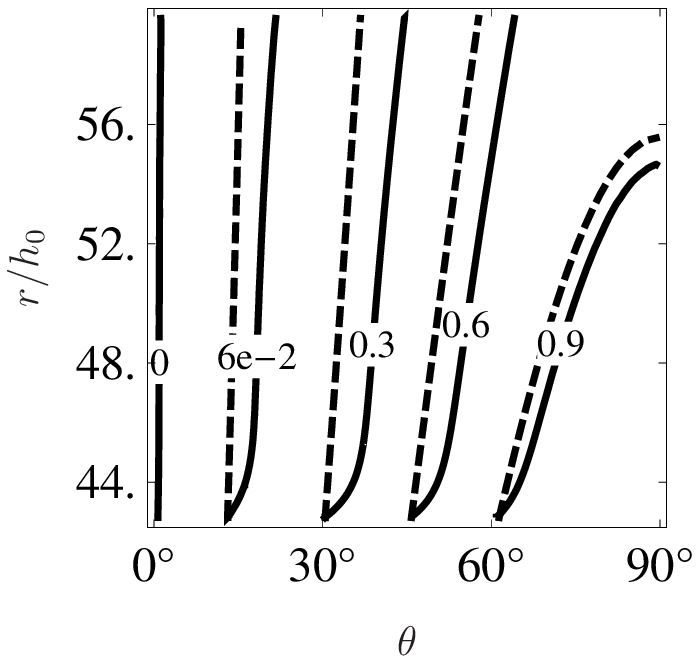}
}
\subfloat[$\probssii$]{\label{fig:flux-sink-surf-1-t21}
\includegraphics[width=0.23\textwidth]{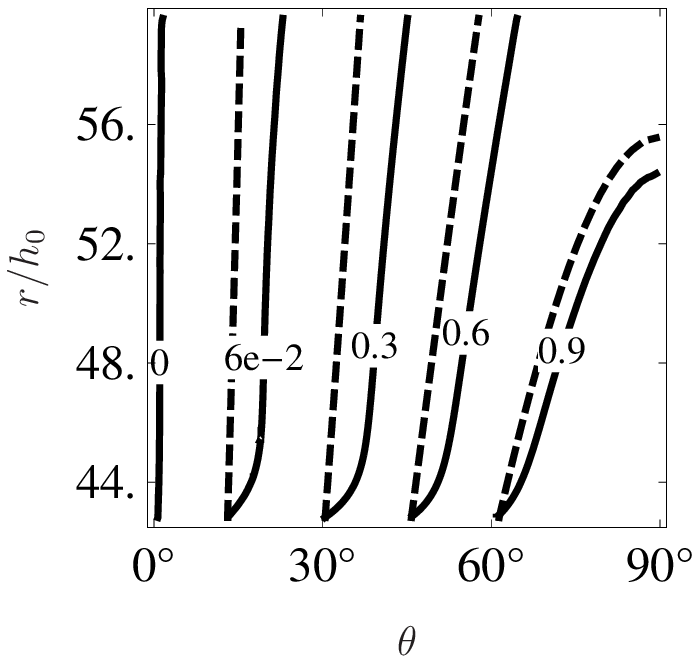}
}
\caption{\label{fig:soft-vs-hard}
Comparison of the hydromagnetic structure of hard- and soft-surface mountains.
Contours of accreted density $\rho \Xmnt / \rhosurf$ (top row) and magnetic flux $\psi / \psistar$ (bottom row) of mountains $\probhii$, $\probsii$, and $\probssii$, at time $t = \Tinj$.
Details are as for Figure~\ref{fig:probhiii}.
Note that the scale of the $r$ axis differs between Figures~\subref{fig:dens-hard-1-t21}--\subref{fig:dens-sink-surf-1-t21}, and between Figures~\subref{fig:flux-hard-1-t21}--\subref{fig:flux-sink-surf-1-t21}.
}
\end{figure*}

\begin{figure}
\centering
\includegraphics[width=0.8\columnwidth]{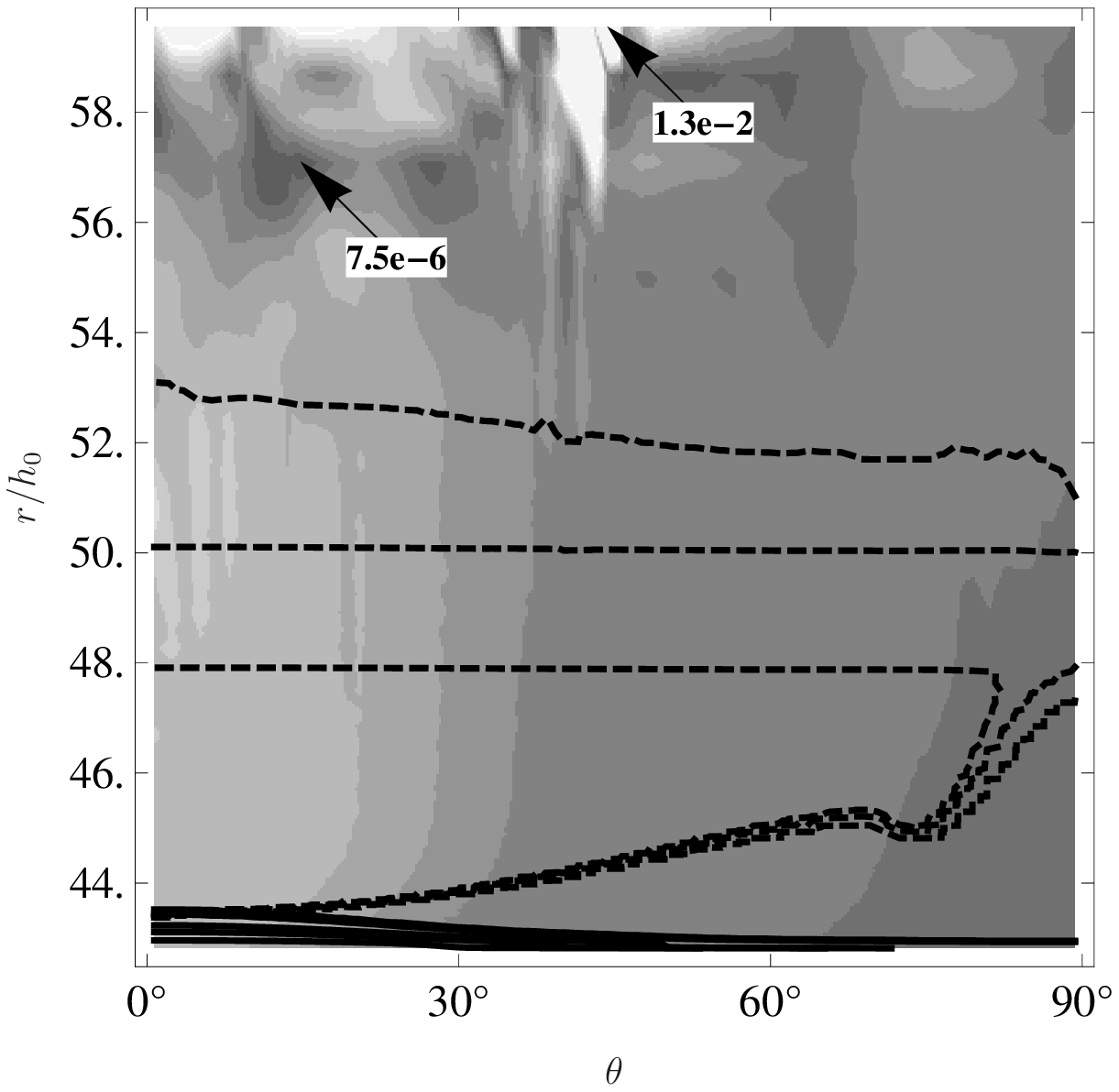}
\caption{\label{fig:diff-soft-vs-hard}
Shaded contours of the absolute, normalised difference in total density $\rho$ between $\probsii$ and $\probssii$, overlaid with the $\rho \Xmnt$ (accreted only) contours of Figures~\subref*{fig:dens-sink-1-t21} and~\subref*{fig:dens-sink-surf-1-t21} (solid and dashed respectively).
The minimum and maximum density differences are indicated with small arrows and labelled in bold.
}
\end{figure}

We next compare the hard-surface equilibrium state, illustrated in Figures~\subref*{fig:dens-hard-2-t21} and~\subref*{fig:flux-hard-2-t21}, with the two experiments where we include sinking.
Figure~\ref{fig:soft-vs-hard} shows contours of $\rho \Xmnt$ and $\psi$ in each of the three scenarios, with $\Ma = \Mii\Mc$.
The hard-surface mountain [Figure~\subref*{fig:dens-hard-1-t21}] spreads appreciably, and the magnetic flux [Figure~\subref*{fig:flux-hard-1-t21}] is significantly displaced towards the equator.
The density contour $\rho\Xmnt/\rhosurf = 10$ begins a distance $\sim 4 h_0 = 215~\mathrm{cm}$ (see Table~\ref{tbl:conversions}) above the injection radius $r = \rmin = \Rstar$ at the pole and sinks below the equator to $\sim 75\%$ of the polar height of the mountain.
In contrast, the same contour of the sinking mountain grown at $r = \rmin$ [Figure~\subref*{fig:dens-sink-1-t21}] begins $\sim 0.6 h_0$ above the injection radius $r = \rmin$ at the pole and sinks below the equator to just $\sim 19\%$ of the polar height of the mountain.
From the $\rho\Xmnt/\rhosurf = 10^{-3}$ contour, we see that the accreted matter is confined to $r - \rmin \lesssim 0.7 h_0$ above the inner boundary at the pole and $r - \rmin \lesssim 0.15 h_0$ at the equator.

In short, the sunk mountain grown at $r = \rmin$ hugs the inner boundary and pole and resembles the initial mass distribution seen in Figure~\subref*{fig:dens-hard-2-t3}.
This is not surprising.
Matter is fed in at $r = \rmin$ with zero velocity, as discussed in section~\ref{ssec:outline}.
It expands outward due to the pressure gradient created as matter piles up at the injection radius; since we are injecting quasistatically, the pressure gradient is small.
On the other hand, the weight of the massive overburden ($\Mbase \approx 10\Ma$) presses down on the added material.
The magnetic flux is displaced [Figure~\subref*{fig:flux-sink-1-t21}], but much less than for the hard mountain.
Field lines remain tied to the inner boundary, bending away in its immediate vicinity (because the slug of injected matter does not rise).
Above this layer, the field lines of the initial and final states remain largely parallel.

The structure of the sunk mountain grown at $r = \Rstar$ [Figure~\subref*{fig:dens-sink-surf-1-t21}] differs from the other two cases.
The contour $\rho\Xmnt/\rhosurf = 10$, tracked above, starts at the pole, remains virtually flat at $\sim 2 h_0$ below the injection radius, bends sharply inward near the equator, moves directly toward the inner boundary, then curves back towards the pole, crossing it again at $\sim 6 h_0$ below the injection radius.
For the previous two mountains, grown from $r = \rmin$, the angular variation in density increases with altitude.
Here the reverse is true: the angular variation density decreases with increasing $r$, up until $r \lesssim \Rstar$, with the greatest variation within $\sim 4 h_0$ of the inner boundary.
While the two sinking scenarios differ in their final distributions of accreted (as opposed to total) density, their final distributions of magnetic flux [Figures~\subref*{fig:flux-sink-1-t21} and~\subref*{fig:flux-sink-surf-1-t21}] are very similar.

The mountain sinks three times further into the fluid base at the pole than at the equator.
This is consistent with how mass is injected according to equation~(\ref{eqn:massfluxrth}); the input flux is $\sim 20$ times greater at the pole than at the equator.
In addition, the magnetic field guides accreted matter sideways as field lines flatten across the surface towards the equator, whereas matter at the pole can sink inward readily along almost vertical flux tubes [e.g. the contour $\psi / \psistar = 6 \times 10^{-2}$ in Figure~\subref*{fig:flux-sink-surf-1-t21}].

The density contours bunch together along the underside of the mountain, spanning five orders of magnitude; the injected matter does not sink below this boundary.
The lowest of the bunched density contours, $\rho\Xmnt/\rhosurf = 10^{-3}$, never reaches the inner boundary; the mountain is floating in isostatic equilibrium with the surrounding fluid base.
The contour rises to only $\sim 1 h_0$ above $\Rstar$ at the equator and $\sim 3 h_0$ at the pole; in contrast it reaches $\Rstar - r \lesssim 6 h_0$ at the pole.
The path of this contour, if overlaid on Figure~\subref*{fig:dens-hard-1-t21}, would trace densities between $2 \times 10^{-3}$ and $50 \rhosurf$.
Finally, note that 0.3\% of the mountain mass is above the stellar surface.
Compared to the other two scenarios, the structure of $\probssii$ is perhaps more reminiscent of an ``iceberg''.

Ultimately we are interested in the final distribution of the total mass, that is, the accreted matter, $\rho \Xmnt$, plus the fluid base it displaces, $\rho(1 - \Xmnt)$.
Does injection at $\rmin$ or $\Rstar$ make a difference?
Figure~\ref{fig:diff-soft-vs-hard} displays the absolute, normalised difference $|\rho_{\rmin} - \rho_{\Rstar}|/|\rho_{\rmin} + \rho_{\Rstar}|$ between the total densities in the two sinking scenarios as a grayscale plot.
The largest differences occur at $r > \Rstar$, where there is little mass, and are therefore unimportant.
For $r < \Rstar$, the difference peaks near the pole but remains less than $\sim 0.8\%$.
In other words, despite the difference in the final distribution of $\rho \Xmnt$ between the two injection scenarios (emphasised by the overlaid contours), the final distribution of $\rho$ is essentially the same.
Injecting at $\rmin$ or $\Rstar$ makes no difference, because the soft base readjusts in each case to yield the same overall equilibrium state.
This is an important result.
It confirms the robustness of the injection method and the argument presented in section~\ref{ssec:outline}: in ideal MHD, the equilibrium state is independent of precisely where matter is initially injected.
In practice, injection at $\rmin$ seems preferable, because it reduces the simulation time (see Table~\ref{tbl:problems}) and improves numerical stability, as illustrated by the failure of run $\probssiiii$.

\subsection{Magnetic line tying}\label{ssec:linetying}

\begin{figure}
\centering
\subfloat[$\probh$ at $r = \rmin = \Rstar$]{\label{fig:angle-hard}
\includegraphics[width=0.49\columnwidth]{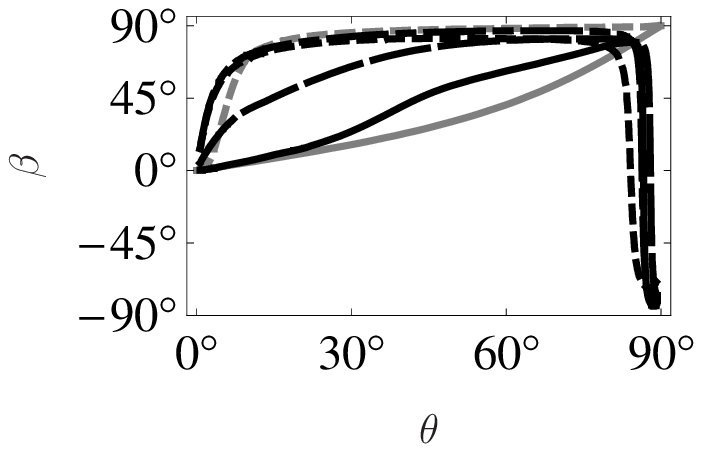}
}
\\
\subfloat[$\probs$ at $r = \Rstar$]{\label{fig:angle-rstar-sink}
\includegraphics[width=0.49\columnwidth]{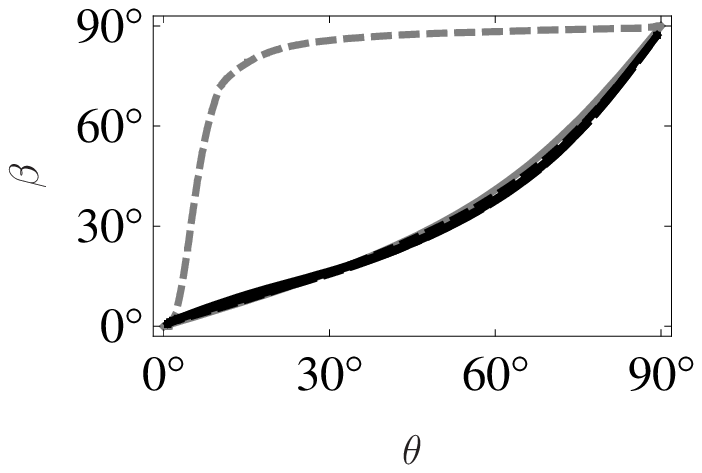}
}
\subfloat[$\probs$ at $r = \rmin$]{\label{fig:angle-rmin-sink}
\includegraphics[width=0.49\columnwidth]{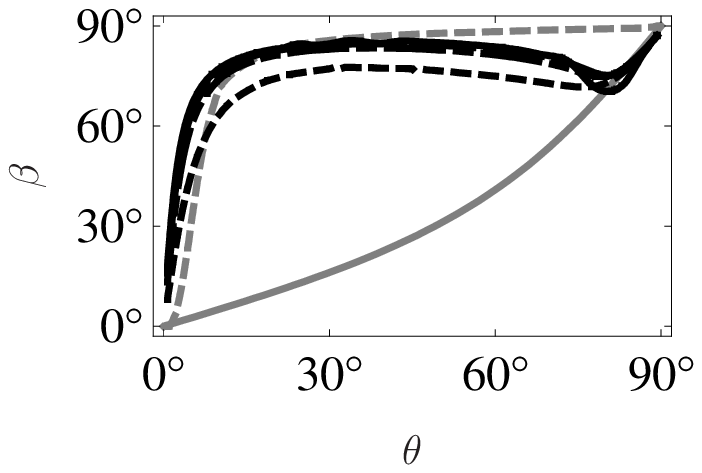}
}
\caption{\label{fig:angle}
Angle between the magnetic field $\Bb$ and the radial unit vector, plotted versus colatitude $\theta$ at~\subref{fig:angle-hard} $r = \rmin = \Rstar$,~\subref{fig:angle-rstar-sink} $r = \Rstar$, and~\subref{fig:angle-rmin-sink} $r = \rmin$, for $\Ma/\Mc = \Mi$ (dotted), $\Mii$ (dot-dashed), $\Miii$ (dashed), and $\Miiii$ (solid).
Results from this paper are plotted in black.
Plotted in gray are $\beta(r,\theta)$ for a \citet{PM04} Grad-Shafranov mountain (dotted), and for a dipole (solid).
}
\end{figure}

Finally, we investigate the assumption of magnetic line tying.
In Figure~\ref{fig:angle} we plot the angle $\beta(r,\theta) = \sin^{-1}(B_{\theta} / |\Bb|)$ between the magnetic field $\Bb$ and the radial unit vector as a function of $\theta$ at $r = \rmin$ and $r = \Rstar$.
We also plot $\beta$ at the inner boundary of a Grad-Shafranov mountain with $\Ma = \Mc$  \citep{PM04}, and $\beta$ for a dipole (independent of radius).
The $\probss$ mountains give the same results as $\probs$.

The hard-surface mountains in Figure~\subref*{fig:angle-hard} behave like the Grad-Shafranov mountain at low $\Ma$ but become increasingly dipolar as $\Ma$ increases.
This is expected; at low $\Ma$, the accreted mass stays close to the pole and distorts the magnetic field there.
As $\Ma$ increases, the mountain spreads over a greater volume, and the magnetic field is distorted less at any particular point.
The sign inversion close to the equator may be caused by numerical reconnection, or by the reflective boundary condition at $\theta = \pi/2$; further tests with a resisitive ideal-MHD solver are needed to make sure.

The soft-surface mountains in Figure~\subref*{fig:angle-rstar-sink} are dipolar at $r = \Rstar$, as expected, but at $r = \rmin$ [Figure~\subref*{fig:angle-rmin-sink}] they more closely resemble the Grad-Shafranov mountain.
One might expect $\beta$ to closely match a dipole at $r = \rmin$, given that the magnetic field lines are tied there, and we choose $\Mbase \gg \Ma$ in order to minimise sideways fluid displacements at the bottom of the soft base.
It is unclear whether the magnetic distortions are artificial, because the injected slug matter enters from below and cannot expand upwards to match accretion from above (see section~\ref{sec:discuss} for further discussion).
Alternatively, kinks in the magnetic field may be communicated rapidly down to arbitrary depths by Alfv\'en waves, even though the Alfv\'en speed $\propto \rho^{-1/2}$ decreases rapidly with depth.
If so, the high breaking strain of the solid, conducting crust \citep{Horowitz-Kadau-2009} assumes even greater importance in enforcing line tying.

We argued, in section~\ref{ssec:outline}, that the final equilibrium state of the mountain is independent of $\rinj$.
In general, a given total $\Ma$ and injected mass flux $\partial \Ma / \partial \psi$ does not define a \emph{unique} ideal-MHD equilibrium.
Matter injected from above spreads sideways faster than it sinks, like a layered cocktail drink, while a slug of matter injected from below forces the base sideways without much movement at the surface [compare Figures~\subref*{fig:dens-sink-surf-1-t21} and~\subref*{fig:dens-sink-1-t21}].
Conceivably, therefore, ZEUS-MP may converge on different equilibria depending on $\rinj$.
The results of section~\ref{ssec:soft-vs-hard} engender confidence that the mountain structure does not depend on $\rinj$; the issue is not definitively settled, however, for the following subtle reason.\footnote{Sterl Phinney, private communication.}

Consider a polar field line in the bottom row of Figure~\ref{fig:probhiii}.
As accretion proceeds, it bends towards the equator until it touches the corner $(r, \theta) = (\rmax, \pi/2)$.
At that point, it instantaneously snaps through some nonzero angle, from $B_r \ne 0$ (free boundary at $r = \rmax$) to $B_r = 0$ (reflecting boundary at $\theta = \pi/2$).
Effectively, this corresponds to a dissipative, reconnection-like event occurring just \emph{outside} the simulation volume, artificially pinching off magnetic loops.\footnote{
The effect can be magnified in ZEUS-MP by increasing the cell size close to $r = \rmax$; eventually ZEUS-MP aborts when $B_r$ diverges close to the $(\rmax, \pi/2)$ corner.}
Such a process is irreversible.
Furthermore, it acts differently on the sequence of quasistatic equilibria that ZEUS-MP hypothetically passes through during slow accretion from above and below, because sideways spreading happens at different altitudes in the two cases.

In the runs presented in this paper, the density in the vicinity of the corner $(\rmax, \pi/2)$ is tiny, as is the mass efflux through the boundary $r = \rmax$ (see section~\ref{ssec:verification}).
One can therefore argue plausibly that the irreversible dissipation at $(\rmax, \pi/2)$, while it exists in principle, does not significantly affect the final state.
There is a chance, however, that if one adds material slowly from above, reconnection (where numerical or real) pinches off one small magnetic loop after another at the equator, as in the Earth's magnetotail.
Resisitive MHD simulations by \citet{VM09R} do not show such behaviour, but they mostly started from preformed Grad-Shafranov equilibria instead of growing the mountain from scratch.
A more careful consideration of this issue is required for future simulations.

\section{Mass quadrupole moment}\label{sec:ellip}

\begin{figure}
\centering
\subfloat[$1 \Mc$]{\label{fig:ellipticity-0}
\includegraphics[width=0.49\columnwidth]{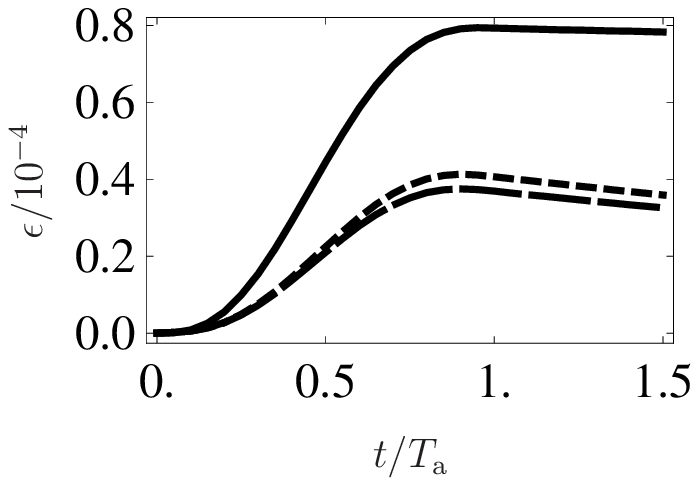}
}
\subfloat[$10 \Mc$]{\label{fig:ellipticity-1}
\includegraphics[width=0.49\columnwidth]{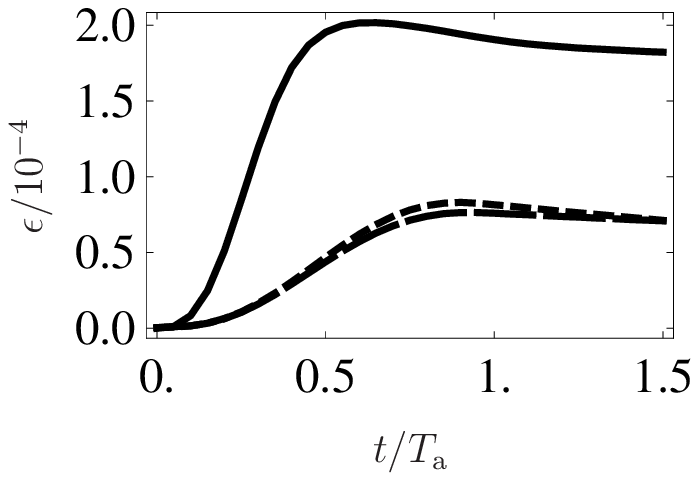}
}
\\
\subfloat[$10^2 \Mc$]{\label{fig:ellipticity-2}
\includegraphics[width=0.49\columnwidth]{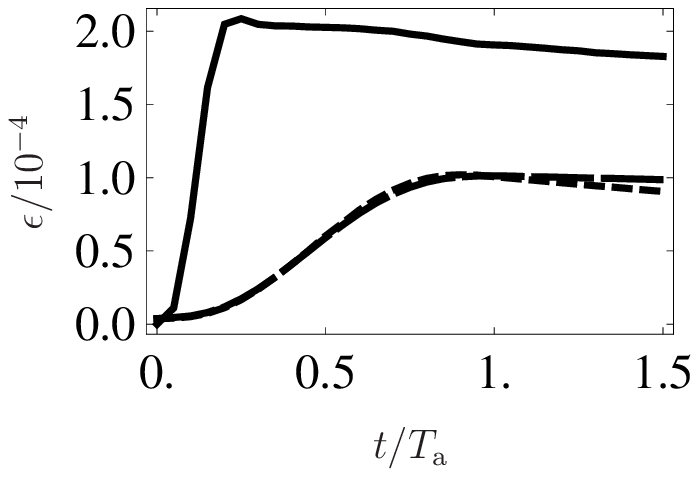}
}
\subfloat[$10^3 \Mc$]{\label{fig:ellipticity-3}
\includegraphics[width=0.49\columnwidth]{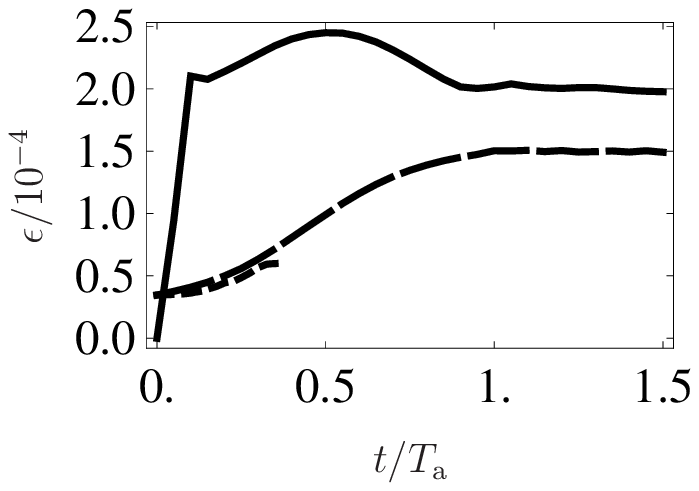}
}
\caption{\label{fig:ellipticity}
Ellipticity as a function of time, for $\Ma / \Mc = \Mi$, $\Mii$, $\Miii$, $\Miiii$ (top left to bottom right), and for mountains $\probh$ (solid), $\probs$ (dashed), and $\probss$ (dotted).
}
\end{figure}

\begin{figure}
\centering
\includegraphics[width=0.8\columnwidth]{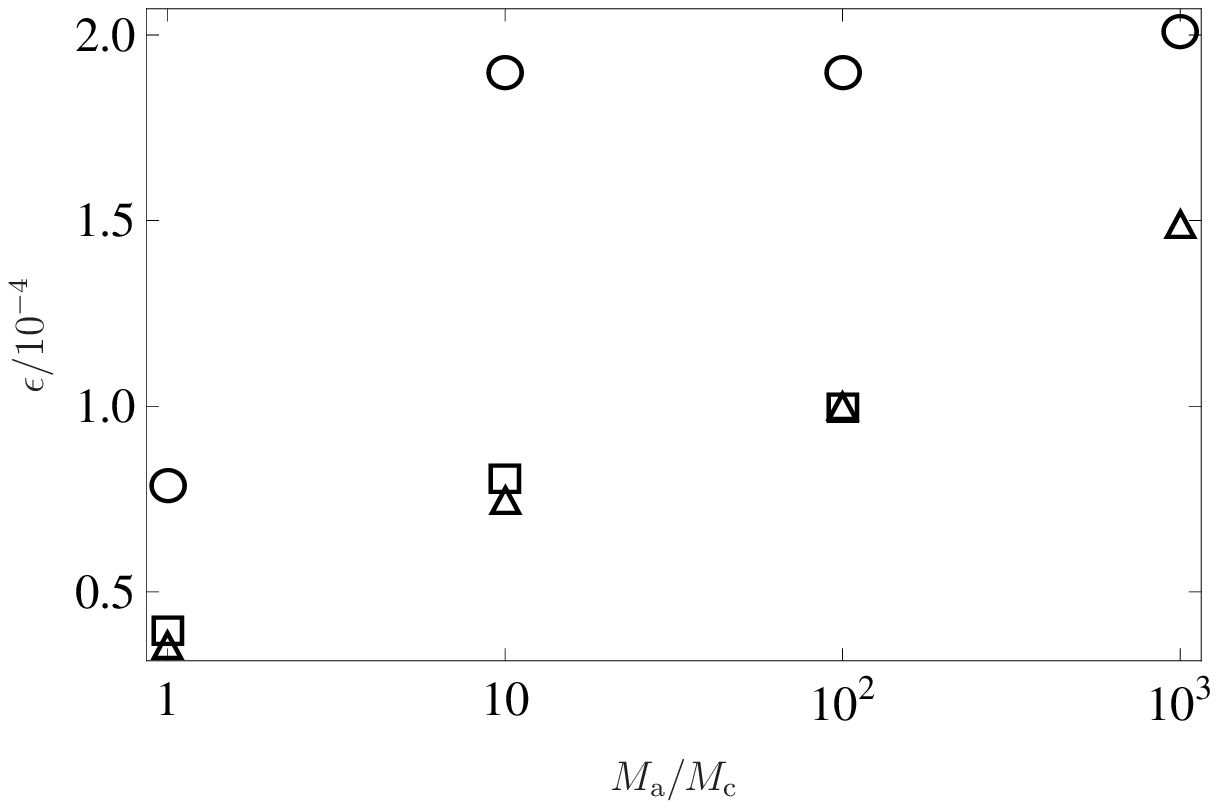}
\caption{\label{fig:ellipticity-Ma}
Ellipticity as a function of $\Ma$, at time $t = \Tinj$, for mountains $\probh$ (circles), $\probs$ (triangles), and $\probss$ (squares).
}
\end{figure}

The distorted hydromagnetic equilibria in section~\ref{sec:compare} have an associated mass quadrupole moment, with principal axis along the pre-accretion magnetic axis, which is quantified in terms of the ellipticity
\begin{equation}\label{eqn:ellipticity}
\epsilon = \frac{ \pi }{ \Izz }
\int_{\rmin}^{\rmax} dr\, r^4
\int_{0}^{\pi/2} d\theta\, \sin\theta (3\cos^2\theta - 1) \, \rho(t,r,\theta) \,,
\end{equation}
with $\Izz = 2 \Mstar \Rstar^2 / 5$.
Figure~\ref{fig:ellipticity} shows $\epsilon$ with respect to time as the mountain grows.
All mountains achieve a nonzero ellipticity at $t = \Tinj$, which decreases negligibly thereafter, confirming the mountains are stable.
The time taken for the hard-surface mountains to converge to their equilibrium values of $\epsilon$ decreases with $\Ma$, from $\sim \Tinj$ ($\Ma = \Mi\Mc$) to $\sim 0.1 \Tinj$ ($\Ma = \Miiii\Mc$); this is consistent with the decreased confinement of the mountain by the magnetic field, i.e. its increased ability to spread.
The ellipticities of the soft-surface mountains for the two injection scenarios are virtually identical; even the incomplete run $\probssiiii$ closely follows $\probsiiii$ up until failure.
As in section~\ref{ssec:soft-vs-hard}, the final density distribution is independent of the injection procedure.

The origin of the uneven behaviour of the ellipticity of $\probhiiii$ [Figure~\subref*{fig:ellipticity-3}] is unknown; we note, however, that its functional dependence on $t$ is similar to that of $\Madot$ [equation~(\ref{eqn:massfluxt}) and Figure~\ref{fig:massflux-t-rth}], and therefore it is likely that the rise and fall of the ellipticity is due to the reconfiguration of the $\probhiiii$ mountain in response to the changing accretion rate.
We note similar undulations in the kinetic energies [Figures~\subref*{fig:energy-sink} and~\subref*{fig:energy-sink-surf}] and, to a lesser degree, in the ellipticity of $\probhii$ [Figure~\subref*{fig:ellipticity-1}].

Figure~\ref{fig:ellipticity-Ma} shows $\epsilon$ (in black) at $t = \Tinj$ as a function of $\Ma$.
The ellipticity of the hard-surface mountains rises by a factor of $\sim 2.5$ as $\Ma$ increases from $\Mi\Mc$ to $\Mii\Mc$ and flattens thereafter, rising by a further 5\% as $\Ma$ increases from $\Mii\Mc$ to $\Miiii\Mc$.
Importantly, accreting further matter does not reduce $\epsilon$; the mountain does not smooth itself out.
The ellipticities of the soft-surface mountains rises by $\sim 60\%$ per decade in $\Ma$.

Figure~\ref{fig:ellipticity-Ma} clearly quantifies the effect of sinking: $\epsilon$ decreases, relative to the hard-surface scenario, by $\sim 50\%$ at $\Ma = \Mi\Mc$, $\sim 60\%$ at $\Ma = \Mii\Mc$, and $\sim 25\%$ at $\Ma = \Miiii\Mc$.

\section{Discussion}\label{sec:discuss}

\begin{table}
\centering
\caption{\label{tbl:parameters}
List of important physical parameters of accreting neutron stars (top part), and a summary of the results of the simulations presented in this paper (bottom part).
}
\begin{tabular}{@{}lll}
\hline

Quantity & Value/Range & Reference \\

\hline

accreted mass & $10^{-4}$ -- $0.8 \Msun$ &
1, 2, 4 \\

accretion timescale & $10^{4}$ -- $10^{6}~\mathrm{yr}$ &
1, 7 \\

density of crust & $10^{9}$ -- $10^{14}~\mathrm{g~cm}^{-3}$ &
5, 10 \\

depth of crust & $\sim 1000~\mathrm{m}$ &
5, 10 \\

initial magnetic field & $10^{12}$ -- $10^{13}~\mathrm{G}$ &
3, 8, 9 \\

temperature & $10^{8}$ -- $10^{9}~\mathrm{K}$ &
6, 10 \\

\hline

ellipticity & $5 \times 10^{-5}$ -- $2 \times 10^{-4}$ &
11 \\

effect of sinking & $\epsilon$ reduced by 25 -- 60 \% &
11 \\

\hline
\end{tabular}
\begin{flushleft}
References:
1.~\citet{Taam-vdHeuvel-1986};
2.~\citet{vdHeuvel-Bitzaraki-1995};
3.~\citet{Hartman-etal-1997};
4.~\citet{Wijers-1997};
5.~\citet{Brown-Bildsten-1998};
6.~\citet{Brown-2000};
7.~\citet{Cumming-etal-01};
8.~\citet{Arzoumanian-etal-2002};
9.~\citet{FaucherGiguere-Kaspi-2006};
10.~\citet{Chamel-Haensel-2008};
11.~This work.
\end{flushleft}
\end{table}

In this paper, we simulate the growth of magnetically confined mountains on an accreting neutron star, with realistic masses $\Ma \lesssim 0.12 \Msun$, under the two scenarios where the mountain sits on a hard surface and sinks into a soft, fluid base.
In the latter scenario, we confirm that the final equilibriun state is independent of the altitude where matter is injected.
We find that the ellipticity of a hard-surface mountain does not increase appreciably for $\Ma \gtrsim \Mii\Mc$, saturating at $\sim 2 \times 10^{-4}$, whereas the ellipticity of a soft-surface mountain continues to increase from $\Ma = \Mii\Mc$ to $\Miiii\Mc$.
Sinking reduces the ellipticity by up to 60\% relative to the hard-surface value.

\citet{Choudhuri-Konar-2002} developed a kinematic model of accretion, which treats sinking in a different (but complementary) way to this paper.
An axisymmetric magnetic field is evolved under the influence of a \emph{prescribed} velocity field, which models the flow of accreted matter from pole to equator, where it submerges and moves towards the core (see their Figure~1).
Ohmic diffusion is included, but, for a subset of the results (where the resistivity $\eta = 0.01$), it is negligible, permitting a direct comparison with this paper.

Figure~5 of \citet{Choudhuri-Konar-2002} shows the evolved configuration of an initially dipolar field which permeates the entire star.
We compare to Figures~\subref*{fig:flux-sink-1-t21} and~\subref*{fig:flux-sink-1-t21} of this paper.
In both models, the magnetic field is distorted significantly by accreted matter spreading towards the equator ($r_\mathrm{m} < r < r_\mathrm{s}$ in \citeauthor{Choudhuri-Konar-2002}; the entire simulation in this paper).
In \citeauthor{Choudhuri-Konar-2002}'s work, the magnetic field is completely submerged beneath the surface and confined to the zone where the submerged accreted matter flows back towards the pole.
In this paper, magnetic field lines still penetrate the surface, implying less effective screening.
Within the core, \citeauthor{Choudhuri-Konar-2002}'s magnetic field remains relatively undisturbed.
Magnetic line-tying is not enforced, but the prescribed radial flow within the core naturally restricts the sideways displacement of the magnetic field there.
If there were sideways motion of the matter within the core, it would modify the degree of magnetic screening, but neither our simulations nor the results of \citeauthor{Choudhuri-Konar-2002} show evidence for such motion.
Extending our simulations deeper into the star to include the core and explore this possibility properly would be a technical challenge; for instance, we would need to incorporate a more realistic equation of state and track even more disparate equilibrium time-scales.

To explain the narrow range in the rotation frequencies of low-mass x-ray binaries \citep{Chakrabarty-etal-2003}, it is proposed that the stars radiate angular moment in gravitational waves at a rate which balances the accretion torque \citep{Wagoner-1984,Bildsten-1998}.
Magnetic mountains are one of a number of physical mechanisms proposed for the associated permanent quadrupole; see \citet{VM09GW} and references therein.
The relationship between $\epsilon$ and the rotation frequency $f$ predicted by torque balance is $f \propto \epsilon^{-2/5}$.
Thus, the 25\% to 60\% reduction in $\epsilon$ due to sinking calculated in this paper increases $f$ by 12\% to 44\%, all other things being equal.
This goes some way towards bringing magnetic mountain ellipticities down to a level consistent with the data, but there is still a long way to go.
Observations to date have found $45 \text{ Hz} < f < 620 \text{ Hz}$ for burst oscillation sources and $182 \text{ Hz} < f < 598 \text{ Hz}$ for accreting millisecond pulsars, implying $6.6 \times 10^{-9} \lesssim \epsilon \lesssim 4.6 \times 10^{-6}$ and $7.2 < 10^{-9} \lesssim \epsilon \lesssim 1.4 \times 10^{-7}$ respectively.
Conversely, the ellipticities of sunk mountains calculated in this paper, $3.5 \times 10^{-5} \lesssim \epsilon \lesssim 1.5 \times 10^{-4}$, imply $11 \text{ Hz} \lesssim f \lesssim 20 \text{ Hz}$.
Clearly, other relaxation mechanisms, like Ohmic diffusion, must also be playing an important role in reducing $\epsilon$, as the observed $f$ require.

The reduction in $\epsilon$ by sinking also reduces the gravitational wave strain \citep[e.g.][]{LSC-CW-Fstat-S2}, $h \propto \epsilon f^2$, by 6\% to 17\%.
This is unlikely, by itself, to rule out the detection of gravitational waves from low-mass x-ray binaries by ground-based interferometric detectors; assuming the signal can be coherently integrated, the loss in $h$ can be compensated for by an increase in the observation time $\propto h^{-2}$ of 13\% to 45\%.
Other difficulties associated with the detection of gravitational waves from low-mass x-ray binaries, such as poorly known orbital parameters and accretion-induced phase wandering \citep{Watts-etal-2008}, are likely to be more important.

\section*{Acknowledgments}

The authors are grateful for supercomputing time allocated on the Australian NCI National Facility [\url{http://nf.nci.org.au}].
KW was supported by an Australian Postgraduate Award.

\bibliography{grow_mountain}

\appendix

\section{Matching a fluid base to a Grad-Shafranov mountain in ZEUS-MP}\label{sec:gsmatch}

We attempted to incorporate a fluid base into the framework of \citet{PM07} and \citet{VM08} in the following ad-hoc manner.
Starting with a Grad-Shafranov equilibrium loaded into ZEUS-MP, we extend the inner simulation boundary, initially at $r = \Rstar$, inwards to create a region $\rmin < r < \Rstar$, containing the fluid base.
The magnetic field $\Bb$ in this region is initialised to a dipole.
At $r = \Rstar$, $B_r$ matches perfectly, but $B_\theta$ is discontinuous [see, e.g., Figure~2 of \citet{PM04}].
The initial density $\rho(t=0,r,\theta)$ is chosen to match the Grad-Shafranov density $\rho_\mathrm{GS}(r,\theta)$ at $r = \Rstar$, and to match an isothermal, non-self-gravitating profile within $r < \Rstar$.
A number of ad-hoc choices of $\rho(t=0,r,\theta)$ were tried, e.g. the maximum of
$\rho_\mathrm{GS}(\Rstar,\theta)$
and
$\rhosurf^\prime \exp [ G \Mstar ( r^{-1} - \Rstar^{-1} ) / \cs^2 ]$,
with
$\rhosurf^\prime = \min_\theta \rho_\mathrm{GS}(\Rstar,\theta)$.
When the combined Grad-Shafranov mountain and fluid base are evolved in ZEUS-MP, the results are undesirable.
Except when $\Rstar - \rmin \ll h_0$, the fluid base is sufficiently far from equilibrium to completely disrupt the Grad-Shafranov mountain, which collapses over a short time-scale $\sim t_0$.

\section{Logarithmic radial grid spacing}\label{sec:logspacing}

The logarithmic grid spacing in $r$ is determined as follows.
The $N_r + 1$ radial cell boundaries $\rmin = r_0,r_1,r_2,\dots,r_{N_r} = \rmax$ are given by $r_{n+1} = r_n + \Delta r_n$, where
\begin{align}
\sum_{n=0}^{N_r-1} \Delta r_n &= \rmax - \rmin \,, \\
\frac{ \Delta r_{n+1} }{ \Delta r_n } &=
\left( \frac{ \Delta r_{N_r - 1} }{ \Delta r_0 } \right)^{1/(N_r - 1)} \,,
\end{align}
and $\Delta r_{N_r - 1} / \Delta r_0$ is the ratio of the maximum to minimum radial grid spacing.
The values of $\Delta r_{N_r - 1} / \Delta r_0$ used in the simulations presented in this paper are given in Table~\ref{tbl:problems}.

\section{Custom injection}\label{sec:extrainj}

We add a new subroutine to ZEUS-MP which is called at the beginning of each time-step $\delta t$.
Within the subroutine, the density $\rho(t,r,\theta)$, mountain concentration $\Xmnt(t,r,\theta)$, and velocity $\vb(t,r,\theta)$ of a grid cell within the injection region (at point $(r, \theta)$ with size $\delta r \times \delta\theta$) are updated, as follows:
\begin{align}
\rho(t + \delta t, r, \theta) &=
\rho(t, r, \theta) + \delta\rho(t,r,\theta) \,,
\\
\Xmnt(t + \delta t, r, \theta) &=
\frac{ \rho(t, r, \theta) \Xmnt(t, r, \theta) + \delta\rho(t,r,\theta) }
{ \rho(t + \delta t, r, \theta) } \,,
\\
\begin{split}
\vb(t + \delta t,r,\theta) &=
\vinj \frac{\Bb(t,r,\theta)}{|\Bb(t,r,\theta)|} \Xmnt(t + \delta t,r,\theta)
\\ &\qquad +
\vb(t,r,\theta) [ 1 - \Xmnt(t + \delta t,r,\theta) ]  \,.
\end{split}
\end{align}
The density increment is given by
\begin{equation}
\delta\rho(t,r,\theta) = \frac{ \Ma }{
2\pi \delta r \delta\theta } \mathcal{I}(t,r,\theta) \,;
\end{equation}
the factor of $2\pi$ comes from the size of the grid cell in the $\phi$ dimension.
The function
\begin{equation}
\mathcal{I}(t,r,\theta) = \frac{ 1 }{ \mathcal{N} }
\int_{t_1}^{t_2} dt
\int_{r_1}^{r_2} dr\, r^2
\int_{\theta}^{\theta + \delta\theta} d\theta\, \sin\theta \,
\frac{ \partial^3 \Ma }{ \partial t \partial r \partial \theta }(t,r,\theta)
\end{equation}
integrates the injected flux given by equation~(\ref{eqn:massflux}); the constant
\begin{equation}
\mathcal{N} =
\int_{0}^{\Tinj} dt
\int_{\rinj}^{\rinj+\drinj} dr\, r^2
\int_{0}^{\pi/2} d\theta\, \sin\theta \,
\frac{ \partial^3 \Ma }{ \partial t \partial r \partial \theta }(t,r,\theta)
\end{equation}
ensures the correct normalisation.
The times
\begin{align}
t_1 &= \min ( t           , \Tinj ) \,, \\
t_2 &= \min ( t + \delta t, \Tinj ) \,,
\end{align}
give the intersection of the current time-step with the injection time interval, and the radii
\begin{align}
r_1 &= \min [ \max ( r           , \rinj ), \rinj + \drinj ] \,, \\
r_2 &= \min [ \max ( r + \delta r, \rinj ), \rinj + \drinj ] \,,
\end{align}
give the intersection of the grid cell with the injection region.

\end{document}